\documentclass[12pt]{article}
\usepackage[utf8]{inputenc}
\usepackage[margin=1in]{geometry}
\usepackage{amsmath}
\usepackage{amsfonts}
\usepackage{amssymb}
\usepackage{algorithm}
\usepackage{blkarray, bigstrut}
\usepackage{algpseudocode}
\usepackage{hyperref}
\usepackage{framed}
\usepackage{caption}
\usepackage{xcolor}
\usepackage[normalem]{ulem}
\usepackage{subcaption}
\usepackage{booktabs}
\usepackage{natbib}
\usepackage{graphicx}
\usepackage[nomarkers, nofiglist, notablist]{endfloat}
\usepackage{verbatim}
\usepackage{bm}

\usepackage{setspace}
\usepackage[pagewise]{lineno}

\usepackage{titlesec}

\newlength{\leftbarwidth}
\newlength{\leftbarsep}
\title{A Bayesian joint longitudinal-survival model with a latent stochastic process for intensive longitudinal data}
\author{Madeline R. Abbott$^{1*}$, Walter H. Dempsey$^1$, Inbal Nahum-Shani$^2$ \\ Lindsey N. Potter$^3$, David W. Wetter$^3$, Cho Y. Lam$^3$, Jeremy M.G. Taylor$^1$ \\
\\
\small $^1$Department of Biostatistics, University of Michigan, \\ \small Ann Arbor, Michigan, U.S.A. \\ \small 
$^2$Institute for Social Research, University of Michigan,\\ \small  Ann Arbor, Michigan, U.S.A. \\ \small 
$^3$Department of Population Health Sciences and Huntsman Cancer Institute, \\ \small  University of Utah, Salt Lake City, UT, U.S.A. \\
\\ \small $^*$ corresponding author, email: mrabbott@umich.edu,\\ \small address: 1415 Washington Heights, Ann Arbor, MI 48109}

\date{  }

\begin{document}

\maketitle

\begin{abstract}
    The availability of mobile health (mHealth) technology has enabled increased collection of intensive longitudinal data (ILD). ILD have potential to capture rapid fluctuations in outcomes that may be associated with changes in the risk of an event. However, existing methods for jointly modeling longitudinal and event-time outcomes are not well-equipped to handle ILD due to the high computational cost. We propose a joint longitudinal and time-to-event model suitable for analyzing ILD. In this model, we summarize a multivariate longitudinal outcome as a smaller number of time-varying latent factors. These latent factors, which are modeled using an Ornstein-Uhlenbeck stochastic process, capture the risk of a time-to-event outcome in a parametric hazard model. We take a Bayesian approach to fit our joint model and conduct simulations to assess its performance. We use it to analyze data from an mHealth study of smoking cessation. We summarize the longitudinal self-reported intensity of nine emotions as the psychological states of positive and negative affect. These time-varying latent states capture the risk of the first smoking lapse after attempted quit. Understanding factors associated with smoking lapse is of keen interest to smoking cessation researchers.
\end{abstract}

\noindent \textit{\textbf{Keywords:}} dynamic factor model, intensive longitudinal data, joint model, mobile health, survival analysis

\noindent \textit{\textbf{Running head:}} Joint model for ILD

\newpage

\titlespacing{\section}{0pt}{\parskip}{-\parskip}
\titlespacing{\subsection}{0pt}{\parskip}{-\parskip}
\titlespacing{\subsubsection}{0pt}{\parskip}{-\parskip}
\setstretch{1.88}

\section{Introduction}

Mobile health (mHealth) technology enables researchers to record longitudinal changes in a variety of biomedical indicators that capture temporal variations in harder-to-measure underlying states. The potentially high frequency of these measurements allows researchers to gain insight into short- and long-term patterns of change in underlying health-related states, such as mood, cognitive function, or disease severity. Here, we use the existing term ``intensive longitudinal data" (ILD) to refer to data consisting of multiple outcomes recorded frequently over time. When these ILD are combined with information on the occurrence of time-to-event outcomes, the ILD can provide insight into factors that elevate the risk of an event. Motivated by ILD and event-time data collected in an mHealth study of smoking cessation, we propose a novel approach for jointly modeling a time-to-event outcome and multiple frequently measured---and possibly rapidly varying---longitudinal outcomes. The key contribution of this work is the development of a joint model suitable for analyzing multivariate ILD. Specifically, we use a multivariate continuous-time stochastic process to (a) flexibly model a smaller number of highly variable latent factors measured through a larger number of longitudinal outcomes and (b) represent risk of a time-to-event outcome by incorporating the latent factors as a time-varying covariates in a hazard model.

\subsection{Related work}

Joint longitudinal-survival models are powerful tools for enabling estimation of the association between temporal variations in longitudinal outcomes and the risk of time-to-event outcomes. A classic joint longitudinal-survival model consists of two parts: a longitudinal submodel and a survival submodel. Joint models attempt to account for the intermittent measurement of the longitudinal outcomes, measurement error, and informative drop-out. For a comprehensive review of joint models, see \cite{tsiatis_2004}. A major challenge to the use of joint models in practice is their computational cost, which rises rapidly as the number of longitudinal outcomes increases. This increasing cost is due to the need to evaluate complex and often intractable integrals across the unobserved random effects in the longitudinal submodel, as well as in the survival function.

Existing literature for jointly modeling multivariate longitudinal outcomes and time-to-event outcomes contains a variety of strategies for dealing with long computation times. These strategies work within both frequentist and Bayesian frameworks. Variations of the two-stage approach—which involves first fitting the longitudinal submodel and then incorporated predicted values (e.g., BLUPS) from the longitudinal submodel as time-varying covariates in the survival submodel—have often been used in settings with multivariate longitudinal outcomes due to the computational speed (e.g., \cite{li_2019}, \cite{signorelli_2021}, \cite{kang_2022}). A well-known drawback of the two-stage approach is the risk of bias in coefficient estimates and so adaptations with bias corrections have also been proposed (e.g., \cite{albert_2010}, \cite*{elmi_2018}, \cite{mauff_2020}). Despite the lower computation time required by the two-stage approach, most existing work has not focused on the ILD setting and so approaches have not been developed specifically for large numbers of longitudinal outcomes.

As an alternative to the two-stage approach, strategies for joint estimation have also been developed for modeling multiple longitudinal outcomes and a time-to-event outcome. Many of these existing approaches have leveraged dimension-reduction tools to help lower computation time. \cite{li_2019} and \cite{li_liu_2021}, for example, proposed joint models in which multiple longitudinal outcomes are summarized using variations of functional principal components analysis (PCA). Factor models, and related approaches such as item response models, have also been used to reduce the dimension of the longitudinal outcomes in the joint model setting; see \cite{he_luo_2016};  \cite{liu_sun_2019}; and \cite*{kang_pan_2022} for example. In these instances, the latent factors that summarize the multiple longitudinal outcomes are then used as time-varying covariates in the survival submodel.

Regardless of whether a dimension-reduction strategy is used to help handle the multiple longitudinal outcomes, a longitudinal submodel must also be specified (either for the latent factors representing summary states or for the observed longitudinal outcomes directly). Simple longitudinal submodels allow for easier integration within the joint estimation framework but with ILD, the larger number of longitudinal measurements allows for specification of a more flexible---and potentially complicated---longitudinal submodel. Numerous spline-based approaches have been developed as a flexible way to model the longitudinal process; for example, \cite{brown_2005}; \cite{muroso_2015}; \cite{rizopoulos_2011}; \cite{kang_pan_2022}; \cite{li_liu_2021}; \cite*{song_2002}; \cite*{tang_2023}; \cite{kang_2022}; \cite*{wong_2022}. Gaussian processes have also been incorporated into the longitudinal submodel to capture important patterns such as serial correlation (e.g., \cite*{proust-lima_2016}; \cite{hickey_2018}).

Although substantial developments have been made in methods for jointly modeling multivariate longitudinal outcomes and survival data, little of this work has focused specifically on the setting of ILD. \cite{rathbun_2013} propose an alternative sampling-based approach for jointly modeling multiple longitudinal outcomes and a time-to-event outcome. Their work is motivated by data collected in a mHealth study and is suitable for analyzing ILD. To deal with the high computational cost of ILD, they avoid specifying a longitudinal submodel altogether through a sampling-based approach. While this approach is computationally fast, the lack of a longitudinal submodel inhibits modeling of measurement error, which we believe is important to account for in our---and many other---settings. More recently, \cite{wong_2022} developed an EM-based approach for non-parametric maximum likelihood estimation of a flexible spline-based joint model. Although this approach was not motivated by ILD, the authors suggest that their method would work with many longitudinal outcomes. This approach, however, does not involve any dimension-reduction of the observed longitudinal outcomes, which are then modeled non-parametrically with splines. In the ILD setting, summarizing the many---and possibly highly correlated---longitudinal outcomes as scientifically meaningful dynamic latent factors has the potential to improve the interpretability of states associated with changes in the risk of an event.

\subsection{Main contributions and outline}

We propose a joint model for ILD that is novel in its specific combination of three submodels. As in existing literature, we take a dimension reduction strategy: rather than using PCA, we use a factor model as it allows more incorporation of scientific understanding into the structure of the model and into the interpretation of the latent factors themselves. Rather than using splines to model the change in multiple latent factors over time, we use a continuous-time multivariate stochastic process; this approach allows us the flexibility to capture abrupt changes in multiple correlated latent factors over time but avoids the complexity of specifying the number and location of knots as in a spline-based approach. We then incorporate the latent factors as time-varying covariates in a hazard regression model. To fit our model, we take a Bayesian approach, which allows us to avoid the need to evaluate complex integrals over a multivariate continuous-time stochastic process. Altogether, this approach enables joint modeling of multivariate ILD and a time-to-event outcome via the novel combination of a dynamic factor model, multivariate stochastic process, and hazard regression model. To the best of our knowledge, the combination of these three submodels with an estimation approach suitable for ILD does not exist in the current literature.

The remainder of this paper is organized as follows: in Section \ref{s:motivating_data}, we briefly introduce the mHealth smoking cessation study motivating this work; in Section \ref{s:method}, we describe our joint model and a corresponding strategy for estimation and inference; in Section \ref{s:sim_study}, we demonstrate the performance of our method via simulation; in Section \ref{s:data_app}, we use our method to analyze data from the smoking cessation study; and in Section \ref{s:discussion}, we provide a discussion.

\section{Motivating data}\label{s:motivating_data}

\begin{figure}
    \centering
    \captionsetup{width=\linewidth}
    \includegraphics[width=\linewidth]{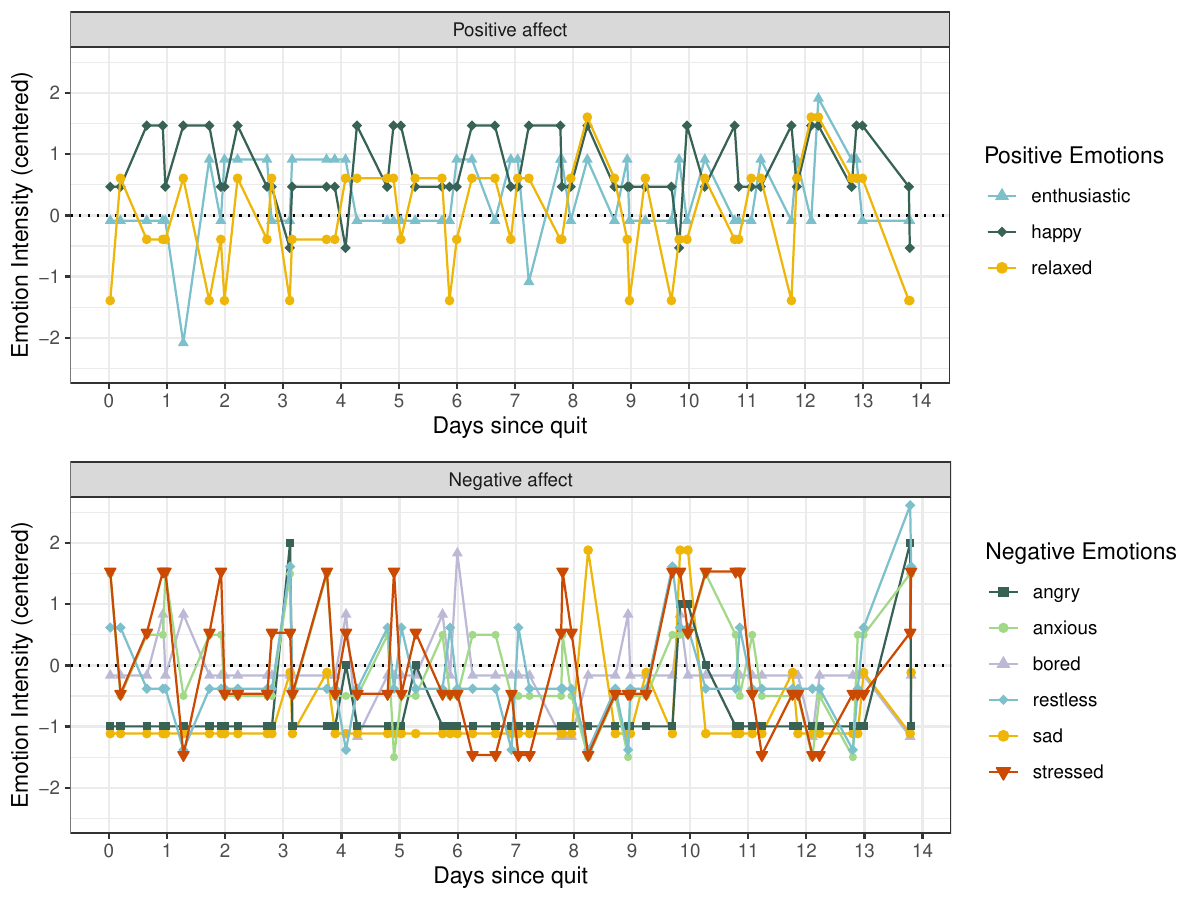}
    \caption{Longitudinal responses to the 9 emotion-related questions for one individual in the smoking cessation study. This individual experienced their first post-quit lapse on day 13.9.} \label{fig:emotions_id3}
\end{figure}

Data motivating this work come from a Houston-based mHealth study. This longitudinal observational cohort study, which ran between 2005 and 2007, followed established smokers for four weeks after they attempted to quit smoking.  This study, called CARE, has been previously described in other publications (e.g., \cite{businelle_2010}, \cite{vinci_2017}).
During the study, the current state and context of individuals were assessed in real time using ecological momentary assessments (EMAs). These EMAs were carried out via surveys sent to mobile Palmtop Personal Computers that prompted individuals to respond to a series of questions capturing their current emotional state and recent cigarette use, among a variety of other social and contextual factors. These EMAs were intended to be sent randomly at four occasions each day. In addition to random EMAs, individuals were instructed to self-initiate EMAs in certain situations (e.g, when feeling a strong urge to smoke, or immediately before or after smoking). We only use information on the longitudinal emotional states reported in the random EMAs. During the four-week post-quit period, individuals responded to an average of 34.3 random EMAs (median = 21.5, range = 1--122). We analyze the 5-point Likert scale responses to the set of nine questions that assessed the current intensity of six negative emotions and three positive emotions over time. The association between smoking and both positive and negative emotions is well-documented in the behavioral science and smoking cessation literature; for example, \cite{vinci_2017} show that positive emotions are associated with a lower likelihood of smoking lapse and \cite{potter_2023} demonstrate that negative emotions are associated with a higher risk of lapse. As such, we aim to use our model to investigate the association between positive and negative affect (a psychological concept related to mood), as captured by the nine emotions measured longitudinally, and the time-to-event outcome of first smoking lapse after attempted quit. Longitudinal responses for one individual are plotted in Figure \ref{fig:emotions_id3}.

We define time--to--lapse as the time until the first episode of cigarette use after attempted quit. To determine the timing of this event, we use information about cigarette use collected from both the random and self-initiated EMAs. Due to uncertainty in the exact time of quit, we restrict our analysis to the subset of individuals who do not report any cigarette use in the first 12 hours after quit (i.e., within 12 hours of the pre-specified 4am quit time on their recorded quit day). Our analytic sample consists of 238 individuals who also responded to the emotion-related questions in at least one random EMA after the first 12 hours of the study.  The time point of 4pm on the recorded quit date serves as time zero in our analysis. In the four weeks of follow-up, 71\% of individuals are observed to have lapsed; the remaining 29\% are censored either at the time of the final EMA to which they responded or at the end of the study. A Kaplan-Meier plot of time-to-lapse is presented in Web Figure 9.

\section{Methods}\label{s:method}

Our proposed joint model consists of three submodels: (i) a measurement submodel, (ii) a structural submodel, and (iii) an event-time submodel. Before describing these models in detail, we first define some notation. Suppose that our data contain information on $i = 1, ..., N$ individuals. For individual $i$, longitudinal outcomes $\bm{Y}_i(t_{ij})$ are measured at occasions $j = 1, ..., n_i$, where $\bm{Y}_i(t_{ij})$ is a vector of length $K$ containing all measurements of the $K$ longitudinal outcomes at time $t_{ij}$. Let $\bm{Y}_i$ be a $(K \times n_i)$-length vector that contains all measurements of the longitudinal outcomes over the $n_i$ occasions. We assume that the longitudinal outcomes are (possibly noisy) observations of a smaller number of $p$ underlying states, represented by $p$-length vector $\bm{\eta}_i(t_{ij})$. We let $T_i$ and $\delta_i$ denote the observed event time and censoring indicator for individual $i$, where $T_i = min(\Tilde{T}_i, C_i), \delta_i = I(\Tilde{T}_i \leq C_i)$ using $\Tilde{T}_i$ as the true event time and $C_i$ as the censoring time.

\subsection{Measurement submodel}\label{ch2:ss:measurement_submod}

To model the set of $K$ longitudinal outcomes observed for individual $i$ at time $t_{ij}$, we use a dynamic factor model. This model is closely related to that developed in \cite{tran_2021} and was also previously presented in \cite{abbott_2023}.  The dynamic factor model is written as $\bm{Y}_i(t_{ij}) = \bm{\Lambda} \bm{\eta}_i(t_{ij}) + \bm{u}_i + \bm{\epsilon}_i(t_{ij})$, where $\bm{\Lambda}$ is a $K \times p$-dimensional loading matrix and $\bm{\eta}_i(t_{ij})$ is $p$-length vector containing the current values of the $p$ latent factors at time $t$, with $p << K$. We make the simplifying assumption that $\bm{\Lambda}$ contains structural zeros and that the location of these structural zeros are known; that is, we assume that we know which of the longitudinal outcomes is a measurement of which of the $p$ latent factors and that each longitudinal outcome measures only a single latent factor. In the motivating study, this assumed structure of $\bm{\Lambda}$ is supported by behavioral science theories that relate certain emotions with certain underlying psychological states. To account for the correlation in repeated measurements, we include $\bm{u}_i \sim N_K(\bm{0}, \bm{\Sigma}_u)$ as an item-specified random intercept. $\bm{\epsilon}_i(t_{ij}) \sim N_K(\bm{0}, \bm{\Sigma}_{\epsilon})$ accounts for measurement error. We assume that $\bm{u}_i$ and $\bm{\epsilon}_i(t_{ij})$ are independent and that $\bm{\Sigma}_{u}$ and $\bm{\Sigma}_{\epsilon}$ are diagonal matrices. We include the random intercept $\bm{u}_i$ to account for differences in underlying levels of the measured longitudinal outcomes across individuals but then use the structural submodel to model correlated change over time.

\subsection{Structural submodel}\label{ch2:ss:structural_submod}

The longitudinal evolution of the $p$ latent factors is assumed to follow a $p$-dimensional multivariate Ornstein-Uhlenbeck (OU) stochastic process. The OU process can be thought of as a continuous-time version of a multivariate autoregressive process and thus is suitable for modeling data with unevenly spaced measurement occasions. We assume that the OU process is stationary with a marginal mean of 0 and is parameterized by two $p \times p$-dimensional matrices, $\bm{\theta}_{OU}$ and $\bm{\sigma}_{OU}$. To ensure a mean-reverting OU process, $\bm{\theta}_{OU}$ is required to have eigenvalues with positive real parts, as discussed in \cite{tran_2021}. $\bm{\sigma}_{OU}$ must have all positive elements. Assuming that the initial value of the latent process, $\bm{\eta}_i(t_{i1})$, is drawn from $N_p(\bm{0}, \bm{V})$, where $\bm{V} = vec^{-1}\big\{(\bm{\theta}_{OU} \oplus \bm{\theta}_{OU})^{-1} vec(\bm{\sigma}_{OU} \bm{\sigma}_{OU}^{\top}) \big\}$, then for $j = 2, ..., n_i$,

\setstretch{1}
\begin{align*}
    \bm{\eta}_i(t_{ij}) | \bm{\eta}_i(t_{i,j-1}) \sim N_p \left(e^{-\bm{\theta}_{OU} (t_{i,j} - t_{i,j-1})} \bm{\eta}_i(t_{i,j-1}), \bm{V} - e^{- \bm{\theta}_{OU} (t_{i,j} - t_{i,j-1})} \bm{V} e^{- \bm{\theta}^{\top}_{OU} (t_{i,j} - t_{i,j-1})}\right)
\end{align*}
\setstretch{1.88}

\noindent Together, the measurement and structural model imply that

\setstretch{1}
\begin{align*} 
    \bm{Y}_{i}(t_{ij}) | \bm{\eta}_i(t_{ij}), \bm{u}_i \sim N_{K}\left( \bm{\Lambda} \bm{\eta}_i(t_{ij}) + \bm{u}_i, \bm{\Sigma}_{\epsilon} \right)
\end{align*}
\setstretch{1.88}

\noindent When describing the joint longitudinal-survival model that will capture the risk of event-time outcomes as a function of the time-varying latent factors, we will refer to the combined structural and measurement submodels as our longitudinal submodel.

\subsection{Survival submodel}

We take a parametric approach to modeling the risk of an event and use the time-varying values of the latent factors to capture vulnerability to the outcome of interest. We define our hazard model as: $h_i \left(t | \mathcal{H}_i(t)\right) = h_0(t; \bm{\gamma}) \exp\big\{f(\mathcal{H}_i(t); \bm{\beta}) + \bm{\alpha} \bm{X}_i\big\}$ where $\mathcal{H}_i(t) = \{\bm{\eta}_i(s), 0 \leq s \leq t \}$ is the history of the latent process up until time $t$, $h_0(t)$ is a parametric baseline hazard function with parameter vector $\bm{\gamma}$, and $\bm{X}_i$ is a vector of baseline covariates with coefficients contained in $\bm{\alpha}$. We write the hazard as a general function of the history of the latent process, $f(\mathcal{H}_i(t); \bm{\beta})$, to allow for flexibility in how the association between the instantaneous risk of an event and the latent process is modeled. For example, we could simply use $f(\mathcal{H}_i(t); \bm{\beta}) = \bm{\beta}^{\top} \bm{\eta}_i(t)$ or we could choose a more complicated function such as $f(\mathcal{H}_i(t); \bm{\beta}) = \bm{\beta}^{\top} \int_s^t \bm{\eta}_i(u) du$.

\subsection{Likelihood}

We make the following assumptions, which are variations of assumptions standard in the joint longitudinal-survival literature: (i) the timing of the longitudinal measurements and censoring is non-informative; (ii) given the random effects and latent factors, the observed longitudinal outcomes and time-to-event outcomes are independent; (iii) conditional on the random effects and latent factors, the observed longitudinal outcomes within an individual are also independent across time; (iv) the latent factors, random effects, and measurement error are independent. Using these assumptions, the joint log-likelihood of our observed data can be written as

\setstretch{1}
\begin{align} \label{eq:llk_with_u}
    log p(\bm{T}, \bm{\delta}, \bm{Y}; \bm{\Theta}) = \sum_{i = 1}^{N} log \left\{ \int p(T_{i}, \delta_{i} | \bm{\eta}_i ; \bm{\Theta}_T) \left[\int p(\bm{Y}_i | \bm{\eta}_i, \bm{u}_i ; \bm{\Theta}_M) p(\bm{u}_i ; \bm{\Theta}_M) d \bm{u}_i \right] p(\bm{\eta}_i ; \bm{\Theta}_{S}) d \bm{\eta}_i \right\}
\end{align}
\setstretch{1.88}

\noindent where $p(T_i, \delta_i | \bm{\eta}_i, \bm{\Theta}_T) = h_i(T_i | \mathcal{H}_i(t))^{\delta_i} exp\left\{ - \int_0^{T_i} h_i(s) ds \right\}$ and $\bm{\Theta} = (\bm{\Theta}_T, \bm{\Theta}_M, \bm{\Theta}_S)$ contains all unknown parameters, with $\bm{\Theta}_T = (\bm{\beta}, \bm{\alpha}, \bm{\gamma})$, $\bm{\Theta}_M = (\bm{\Lambda}, \bm{\Sigma}_u, \bm{\Sigma}_{\epsilon})$, and $\bm{\Theta}_S = (\bm{\theta}_{OU}, \bm{\sigma}_{OU})$.

The main challenges to fitting our model stem from two integrals in the likelihood: one over the multivariate OU stochastic process and another within the survival function. For the integral over the multivariate OU process, we could use Monte Carlo integration or the fully exponential Laplace approximation \citep*{rizopoulos_2009}; instead, we opt for a fully Bayesian approach. We use Hamiltonian Monte Carlo (HMC) sampling as implemented in the software Stan \citep{carpenter_2017}. In the following paragraphs, we address additional challenges stemming from the large number of latent variables in our model, identification of the longitudinal submodel parameters, and calculation of the survival function.

\paragraph{Reducing the number of latent variables:} We could write the distribution of the observed longitudinal outcome conditional on all latent variables (i.e., the latent factors and the random intercept) as in Equation \ref{eq:llk_with_u}.  However, attempting to estimate so many latent parameters within Stan is challenging and we found that using the likelihood in Equation \ref{eq:llk_with_u} resulted in Monte Carlo samples with very poor mixing and high computational cost. With this parameterization of the likelihood, the number of parameters also increases by $N+2$ for each additional longitudinal outcome, which is potentially problematic in the ILD setting. To reduce the number of parameters that we need to sample, we can instead integrate the distribution of the observed longitudinal outcome over the random intercept so that the likelihood is conditional only on the latent factors. That is, the longitudinal component of the likelihood in our joint model is $\bm{Y}_i | \bm{\eta}_i$, rather than $\bm{Y}_i | \bm{\eta}_i, \bm{u}_i$. As a result, our posterior distribution becomes

\setstretch{1}
\begin{align} 
    p(\bm{\eta}_i, \bm{\Theta}) \propto  \prod_{i = 1}^{N} p(T_i, \delta_i | \bm{\eta}_i, \bm{\Theta}) p(\bm{Y}_i | \bm{\eta}_i, \bm{\Theta}) p(\bm{\eta}_i | \bm{\Theta}) p(\bm{\Theta})
\end{align}
\setstretch{1.88}

\noindent where $\bm{Y}_i | \bm{\eta}_i \sim N_{K\times n_i}\left((\bm{I}_{n_i} \otimes \bm{\Lambda})\bm{\eta}_i, (\bm{J}_{n_i} \otimes \bm{\Sigma}_u) + (\bm{I}_{n_i} \otimes \bm{\Sigma}_{\epsilon})\right)$, $\otimes$ is a Kronecker product, $\bm{J}_{n_i}$ is a $n_i$-dimensional matrix of ones, and $\bm{I}_{n_i}$ is a $n_i$-dimensional identity matrix. This posterior distribution now only involves the covariance matrix of the random intercept, $\bm{\Sigma}_u$, rather than the random intercepts themselves.

\paragraph{Identifying the longitudinal submodel parameters:} Because we use a dynamic factor model as our longitudinal submodel, we require additional assumptions to identify both the loadings matrix $\bm{\Lambda}$ and the structural submodel parameters $\bm{\theta}_{OU}$ and $\bm{\sigma}_{OU}$. Common approaches to identifiability of factor models include either fixing the scale of the loadings matrix or fixing the scale of the latent factors; here, we fix the scale of the latent factors by modeling them on the correlation scale. In \cite{tran_2021}, the authors incorporated an OU process into a dynamic factor model and, rather than directly estimating $\bm{\theta}_{OU}$ and $\bm{\sigma}_{OU}$, they estimated $\bm{\theta}_{OU}$ and the stationary correlation matrix of the OU process, $\bm{V}$. We take the same approach here and parameterize our OU process in terms of $\bm{\theta}_{OU}$ and $\bm{\rho}$, where $\bm{\rho}$ is vector of unknown parameters corresponding to the off-diagonals of $\bm{V}$. 
Converting between the $(\bm{\theta}_{OU}, \bm{\sigma}_{OU})$ and $(\bm{\theta}_{OU}, \bm{\rho})$ parameterizations of the OU process is straightforward (see Web Appendix A.1).

\paragraph{Calculating the survival function:}  The final challenge to fitting our model involves calculating the survival function. In our likelihood, evaluating the term corresponding to the survival submodel, $p(T_i, \delta_i | \bm{\eta}_i; \bm{\Theta})$, requires integrating over the hazard function, which depends on values of the latent factors at all times from 0 to $T_i$. We approximate this integral using a sum across small but discrete time intervals via a midpoint rule. For example, consider a simple hazard model that depends on a constant baseline hazard and the current values of two latent factors: $h_i(t | \mathcal{H}_i(t)) = exp\{\beta_0 + \beta_1 \eta_{1i}(t) + \beta_2 \eta_{2i}(t) \}$. Then, we approximate the survival function as

\setstretch{1}
\begin{align} 
    &p(T_i, \delta_i | \mathcal{H}_i(t)) = h_i(T_i | \mathcal{H}_i(t))^{\delta_i} exp\left\{ - \int_0^{T_i} h_i(s) ds \right\} \label{eq:cumul_haz}\\
    &\approx  h_i(T_i | \mathcal{H}_i(t) )^{\delta_i} exp\left\{ -\sum_{m=1}^{M_i} \frac{1}{2}[h_i(s_{m-1}) + h_i(s_{m})] (s_{m} - s_{m-1})  \right\} 
\end{align}
\setstretch{1.88}

\noindent where $s_m, m = 1, ..., M_i$ correspond to times on a fine grid of $M_i$ points going from $s_0 = 0$ to $s_{M_i} = T_i$. Note that the integral in Equation \ref{eq:cumul_haz} would be straightforward to evaluate if the latent factors were modeled using a mixed model with a linear term for time. Because we use a continuous time OU stochastic process to model the evolution of the latent factors, we must integrate over a complicated function of time and so we use this midpoint approach to approximate the integral instead. In practice, this grid is made up of both measurement times and additional grid points. We discuss how to determine the density of this grid and how to distribute each individual's set of $M_i$ points from 0 to $T_i$ later in Section \ref{ss:grid}.

\section{Simulation study}\label{s:sim_study}

To investigate the empirical performance of our proposed method, we assess the bias of point estimates and coverage of credible intervals via simulation. We use the design of the mHealth smoking cessation study described in Section \ref{s:motivating_data} to inform our simulation study. We set the sample size of a single simulated dataset to $N = 200$ individuals. We assume that $K = 4$ longitudinal outcomes are measured repeatedly over time, where the maximum follow-up time is 28 days and the specific pattern of measurements varies across four difference scenarios. For each of the four measurement scenarios, we generate data under two different sets of true parameters (called setting 1 and setting 2). Setting 1 corresponds to a true OU process with higher correlation and setting 2 corresponds to a true OU process with lower correlation. All data-generating parameters are given in Web Appendix B.2.

All individuals are assumed to have one measurement occasion at baseline. We assume that the $K = 4$ observed longitudinal outcomes, $\bm{Y}_i$, are measurements of two latent factors, $\bm{\eta}_1$ and $\bm{\eta}_2$, where $\bm{Y}_i(t) \sim N_K\left(\bm{\Lambda} \bm{\eta}_i(t) + \bm{u}_i, \bm{\Sigma}_{\epsilon}\right)$. Our placement of the structural zeros within $\bm{\Lambda}$ means that $\bm{Y}_1$ and $\bm{Y}_2$ are measurements of $\bm{\eta}_1$ and that $\bm{Y}_3$ and $\bm{Y}_4$ are measurements of $\bm{\eta}_2$. We assume that the true hazard model that underlies our observed events is $h_i(t) = \exp\big\{ \beta_0 + \beta_1 \eta_{1i}(t) + \beta_2 \eta_{2i}(t) \big\}$.

The number of times that the longitudinal outcomes are observed varies by setting (i.e., true parameter values) and by measurement pattern, as summarized below.

\begin{itemize}
    \item[] \textbf{Pattern 1}: Measurements occur frequently and with constant probability. Individuals have an average of 19 and 25 longitudinal measurements in setting 1 and 2, respectively.
    \item[] \textbf{Pattern 2}: Measurements occur less frequently but still with constant probability. Individuals have an average of 6 and 5 longitudinal measurements in setting 1 and 2, respectively.
    \item[] \textbf{Pattern 3}: Measurements are distributed according to the measurement times bootstrapped from the motivating mHealth study, CARE. Individuals have an average of 34 and 38 longitudinal measurements in setting 1 and 2, respectively.
    \item[] \textbf{Pattern 4}: Measurements are clustered together and distributed according to probabilities following a truncated cosine function of time. Individuals have an average of 30 and 21 longitudinal measurements in setting 1 and 2, respectively.
\end{itemize}

Across the 100 simulated datasets in each setting with measurement patterns 1, 2, and 4, the observed event rates are, on average, 75\% in setting 1 and 71\% in setting 2. Simulated datasets with measurement pattern 3 have a slightly higher observed event rate: the average in setting 1 is 85\% and in setting 2 is 81\%.

\subsection{Discrete approximation of the survival function}\label{ss:grid}
Recall that the midpoint rule for evaluating the cumulative hazard function requires defining a grid of $M_i$ points from 0 to $T_i$ for each individual. This grid can vary in both the density and the distribution of the points. A finer grid corresponds to a more accurate approximation of the cumulative hazard but requires increased computation time; on the other hand, a coarser grid potentially decreases the accuracy of this approximation but is less computationally intensive. In our simulation study, we consider various grid densities (i.e., grids that vary in the average gap in time between grid points). We also consider strategically placing the grid points with increased density in areas where the (estimated or true) hazard is higher. Through simulations, we find that the posterior distributions of our parameters are not sensitive to the distribution of the grid points (i.e., we placed the grid points closer together where the \textit{true} hazard function is higher) and so we opt to take the simpler approach of placing these grid points at equally spaced intervals. In the following simulations, we vary the width of the grid between 0.2, 0.8, and 1.2 days. These grid widths are used to define the spacing of additional points that are added to the longitudinal measurement times; together, these added points and the longitudinal measurement times make up the $M_i$ points used to approximate the survival function. When defining the grid, we require that grid points are specified at all event/censoring times but drop any other grid points that are too close to the measurement occasions, where too close is defined as within 30\% of the specified grid distance (e.g., for a grid of 1.2, any added grid points within 0.36 units of time of a measurement occasion would be dropped). In our simulation study, we also consider a scenario in which we only add grid points at event/censoring times and not at intermediate time points.

\subsection{Simulation results}\label{ss:sim_results}
\begin{figure}
    \centering
    \captionsetup{width=\linewidth}
    \includegraphics[width=\linewidth]{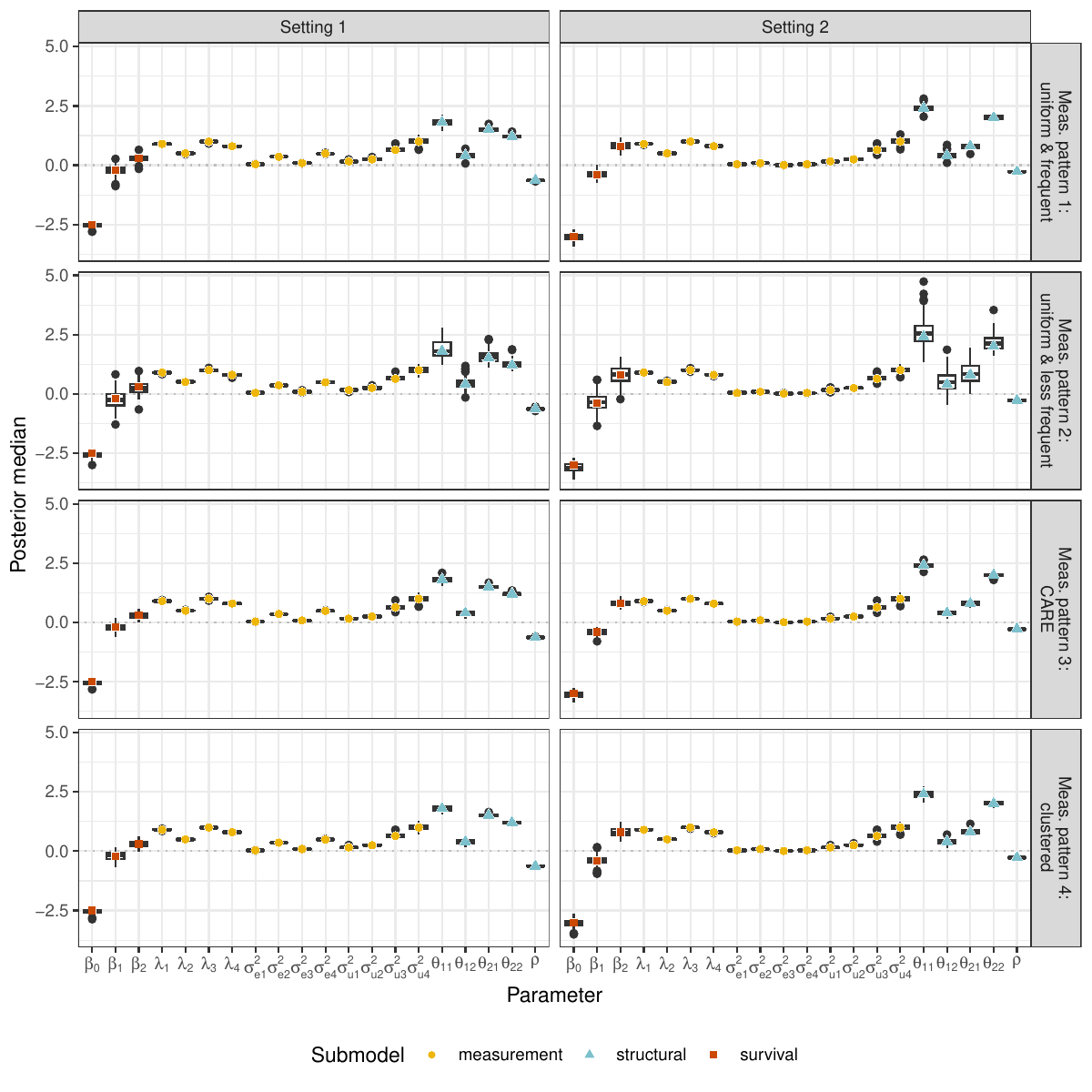}
    \caption{For data generated under settings 1 and 2 with each of the four measurement patterns, we use box plots to summarize the distribution of the \textbf{posterior medians for all parameters} across the 100 simulated datasets. When fitting the model, we assume a \textbf{grid width of 0.8}. True parameter values are indicated with colored dots.} \label{fig:post_medians_grid0.8}
\end{figure}

\begin{figure}
    \centering
    \captionsetup{width=\linewidth}
    \includegraphics[width=\linewidth]{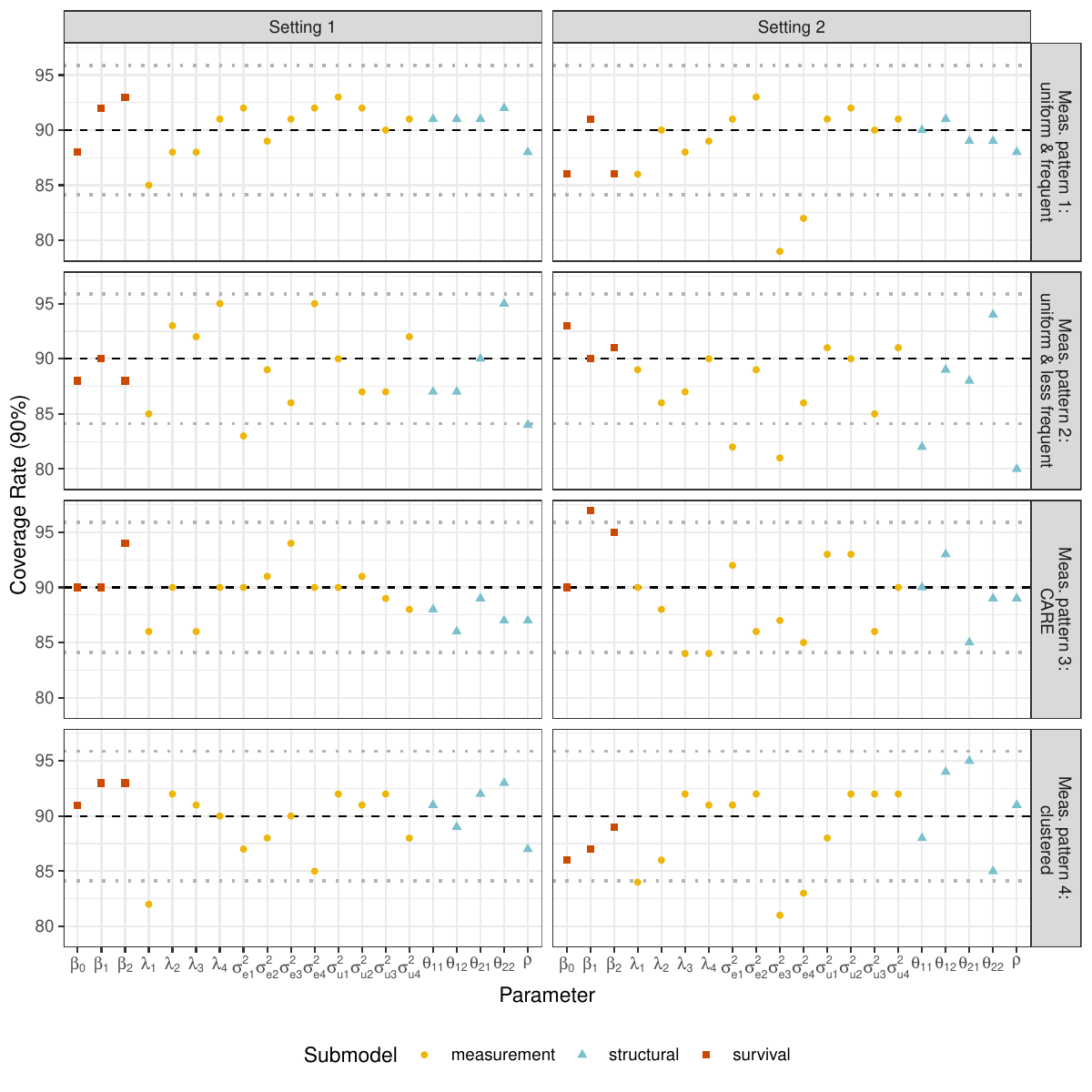}
    \caption{For data generated under settings 1 and 2 with each of the four measurement patterns, we summarize the \textbf{coverage rate of 90\% credible intervals} across the 100 simulated datasets with the colored dots. The black horizontal dashed lines indicate target coverage and the dotted grey lines corresponds to the upper and lower bounds of a 90\% binomial proportion confidence interval for a probability of 0.9.} \label{fig:cov_rate_grid0.8}
\end{figure}

For each simulated dataset generated under setting 1 and 2 with measurement patterns 1-4, we run the HMC sampler using 1 chain for 3,000 iterations and discard the first 2,000 iterations as burn-in. The sampler allows the user to specify initial parameter estimates; we specify reasonable initial values that have the correct sign and approximately correct order of magnitude. Exact initial values, along with prior distributions, are given in Web Appendix B.3 and B.4. To ensure that the OU process in our structural submodel is mean-reverting, we implement the constraints on $\bm{\theta}_{OU}$ that are derived in \cite{tran_2021} and summarized here in Web Appendix A.2. We assess convergence via trace plots and find satisfactory mixing. To summarize our point estimates, we present the distribution of the posterior medians across the 100 simulated datasets in each setting and measurement scenario in Figure \ref{fig:post_medians_grid0.8}, assuming a grid width of 0.8 when fitting the model. To assess the coverage of the 90\% credible intervals, we summarize the average coverage rate for each parameter in Figure \ref{fig:cov_rate_grid0.8}. As the number of added grid points increases, computation time increases substantially (see Web Figure 5).

\begin{figure}
    \centering
    \captionsetup{width=\linewidth}
    \includegraphics[width=0.75\linewidth]{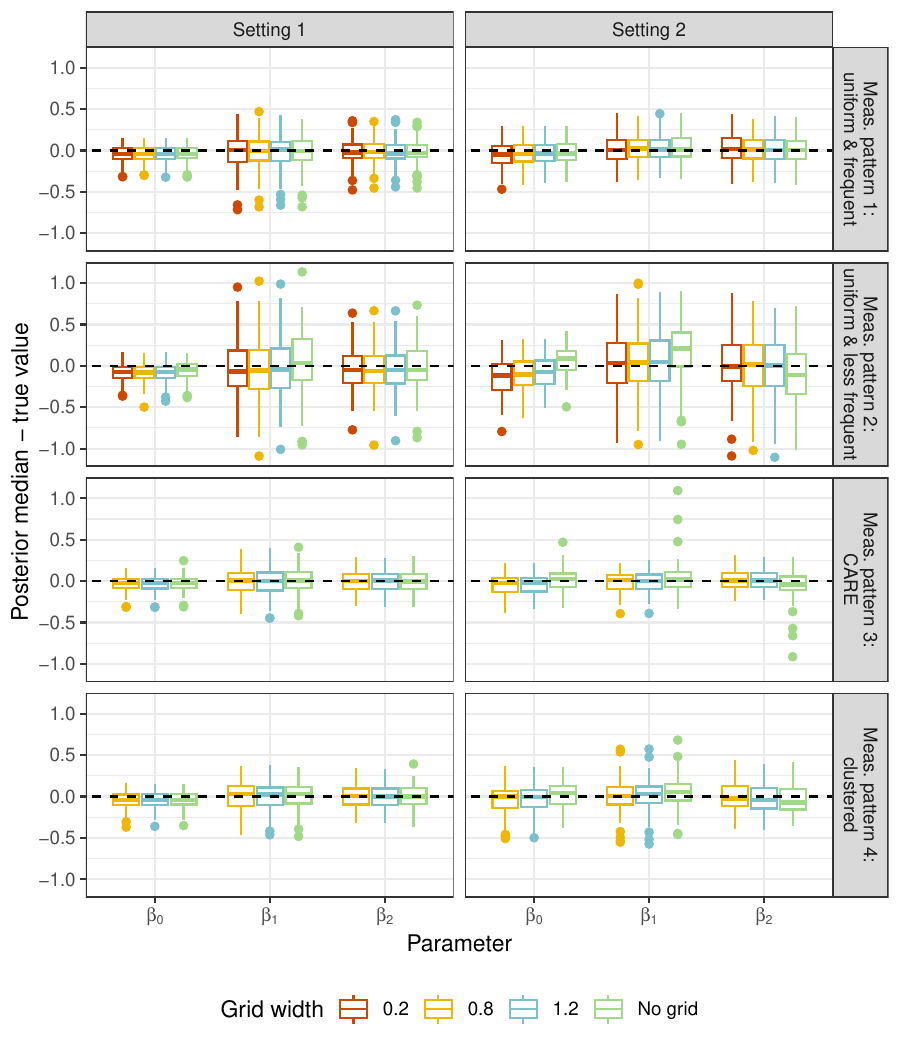}
    \caption{For data generated under settings 1 and 2 with each of the four measurement patterns, we use box plots to summarize the distribution of the \textbf{difference between the posterior medians and true values by grid width for the survival submodel parameters} across the 100 simulated datasets. We fit the model using a grid of 0.2, 0.8, 1.2, and no grid; we do not use the finest grid when fitting the model to data generated under measurement patterns 3 and 4 due the high computation times and lack of bias using the coarser grids.} \label{fig:post_medians_survival}
\end{figure}

We find that our approach, which uses the discrete approximation of the survival function, recovers unbiased estimates of the parameters and returns posterior distributions of appropriate width. We also find that in our setting of ILD, the point estimates and credible intervals are not particularly sensitive to the choice of grid density (see Web Figure 6 and 8 for summaries of posterior medians and coverage rates across all grid widths).  Given that grid points would not be needed if we were to fit only the longitudinal submodels (i.e., jointly fit the measurement and structural submodels), we do not expect these parameter estimates to be sensitive to the grid width. In a few instances in which the measurement occasions are irregular, however, adding grid points can help with convergence of the longitudinal submodel parameters; see Web Appendix C for more discussion.  For the survival submodel parameters, adding grid points at intermediate time points and not only at the event/censoring times does appear to slightly improve our estimates when the longitudinal measurement occasions are infrequent and the correlation of the true OU process decays quickly (i.e., setting 2 under measurement pattern 2), as shown in Figure \ref{fig:post_medians_survival}.

\section{Analysis of smoking cessation data}\label{s:data_app}

We illustrate our method by using it to jointly model the self-reported intensity of nine different emotions recorded longitudinally and the instantaneous risk of a lapse in smoking cessation after attempted quit. We assume that the three positive emotions---enthusiastic, happy, and relaxed---are measurements of the latent psychological state of positive affect and that six negative emotions---sad, angry, anxious, restless, stressed, and bored---are measurements of the latent psychological state of negative affect. We then model time until first lapse as a function of the current values of positive and negative affect. We also adjust for two baseline covariates: pre-quit smoking history and partner status. Pre-quit smoking history is defined here as a binary variable based on the average number of cigarettes smoked per day, where more than 20 cigarettes/day corresponds to heavy smoking. We adjust for this baseline covariate because tobacco dependence is likely associated with the risk of lapse after attempted quit. Prior studies have found positive associations between partner involvement and outcomes of smoking cessation attempts (e.g., \cite*{britton_2019}) and so we also adjust for partner status here in our survival submodel. 

Our survival submodel is $h_i(t) = h_0(t) exp\{ \beta_1 \eta_{1i}(t) + \beta_2 \eta_{2i}(t) + \alpha_1 D_i + \alpha_2 P_i \}$. We specify a flexible piecewise constant baseline hazard for $h_0(t)$; $\eta_{1i}(t)$ and $\eta_{2i}(t)$ are the time-varying latent factors interpreted as positive affect and negative affect, respectively; $D_i$ is the baseline measure of pre-quit smoking history (1 = 20 or more cigarettes per day, 0 = less than 20 cigarettes per day); and $P_i$ is an indicator variable for partner status (1 = lives with a partner or spouse, 0 = everyone else). More details on the specification of this baseline hazard, along with priors, are given in Web Appendix D.1 and D.2.

We initialize parameter estimates using a two-stage approach: we first fit only the longitudinal submodel (via Stan) and use posterior samples of the latent process to fit the hazard regression model (via \texttt{flexsurv} \citep{flexsurv}). For simplicity, we assume a constant (exponential) baseline hazard during initialization. Posterior medians---for the longitudinal submodel parameters---and maximum likelihood estimates---for survival submodel parameters---are used as initial parameter values for joint estimation. To fit the joint model, we run the HMC sampler with 4 chains for 4,000 iterations and discard the first 3,000 samples as burn-in. We assess mixing via trace plots (see Web Figure 10). We also considered a survival submodel with a Weibull baseline hazard, but after comparing the goodness-of-fit of these two joint models via the distribution of predicted survival probabilities, we concluded that the piecewise constant baseline hazard better fit our data. More details on our approach to assessing goodness-of-fit are given in Web Appendix D.3. We present results for the joint model with the piecewise constant baseline hazard below. 

\begin{figure}
    \centering
    \captionsetup{width=\linewidth}
    \includegraphics[width=\linewidth]{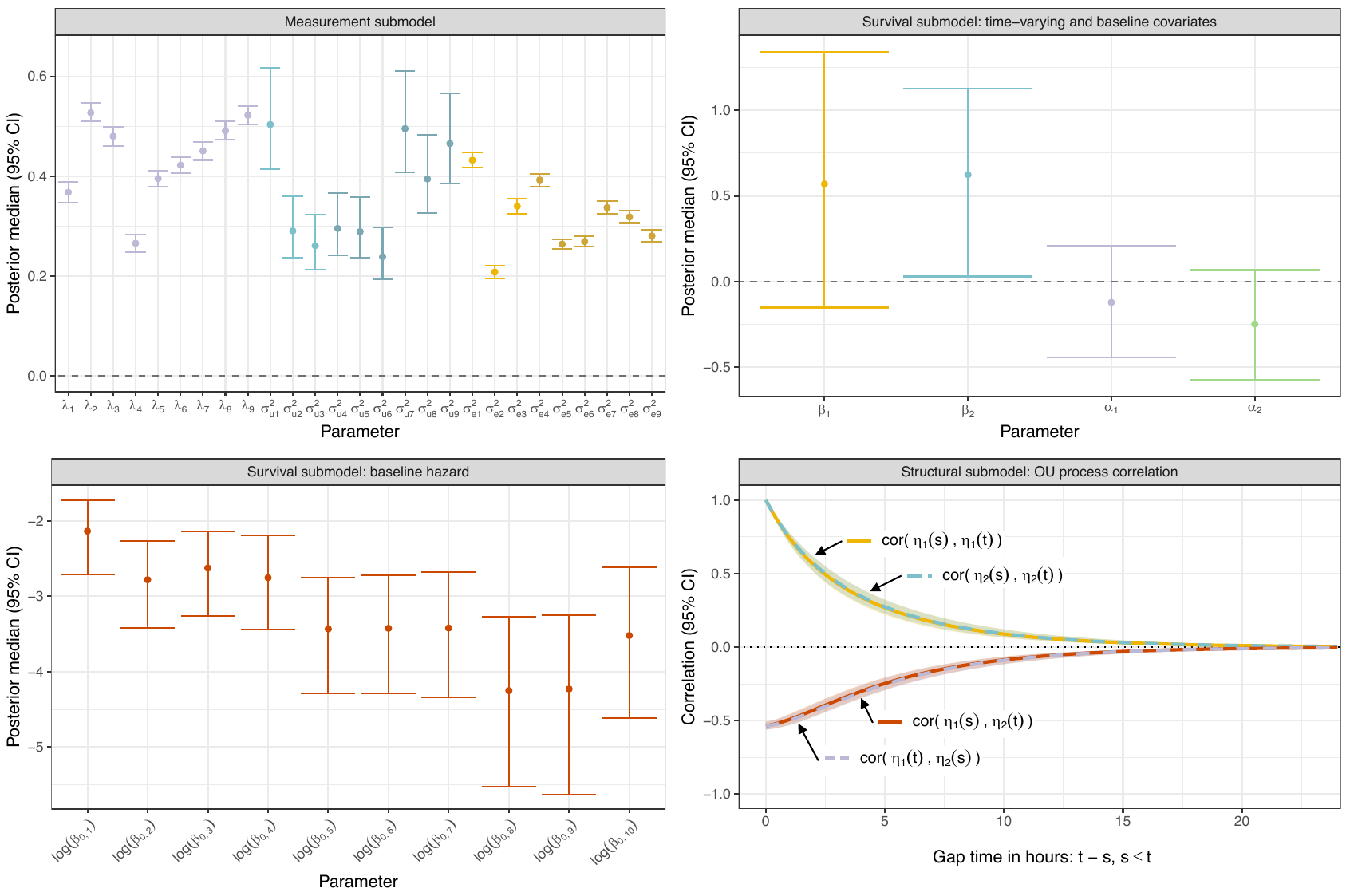}
    \caption{Plot of posterior medians and 95\% credible intervals for parameters in the joint model with a piecewise constant baseline hazard fit to data from the mHealth smoking cessation study. Structural submodel parameters are presented via the estimated correlation decay in latent factors across increasing time intervals (see Web Appendix D.4 for more details on the construction of this subplot). Subscripts index the measured emotions as: 1 = enthusiastic, 2 = happy, 3 = relaxed, 4 = bored, 5 = sad, 6 = angry, 7 = anxious, 8 = restless, 9 = stressed. $\eta_1(t)$ is interpreted as positive affect and $\eta_2(t)$ is interpreted as negative affect.} \label{fig:care_forest_plot}
\end{figure}

In Figure \ref{fig:care_forest_plot}, we plot posterior medians and 95\% credible intervals for the parameters in each submodel. From our structural submodel, we see that our two latent factors representing positive affect ($\eta_1$) and negative affect ($\eta_2$) have a negative correlation of approximately -0.54. We also find that the posterior estimates of the parameters in the structural submodel show fairly symmetric behavior across both positive and negative affect; that is, the correlation shows similar patterns of decay as positive and negative affect are measured across increasing intervals of time. From the measurement submodel, we find that measurements of happy have the largest loading onto the latent factor representing positive affect, measurements of stressed have the largest loading onto the latent factor representing negative affect, and measurements of bored have the smallest loading onto negative affect. Finally, from the survival submodel, we find that a one-standard deviation increase in negative affect is associated with a 1.87-times increase (95\% CI: 1.03-3.10) in the hazard of a lapse. Neither of our baseline covariates are significantly associated with changes in the hazard of lapse.

\begin{figure}
    \centering
    \captionsetup{width=\linewidth}
    \includegraphics[width=\linewidth]{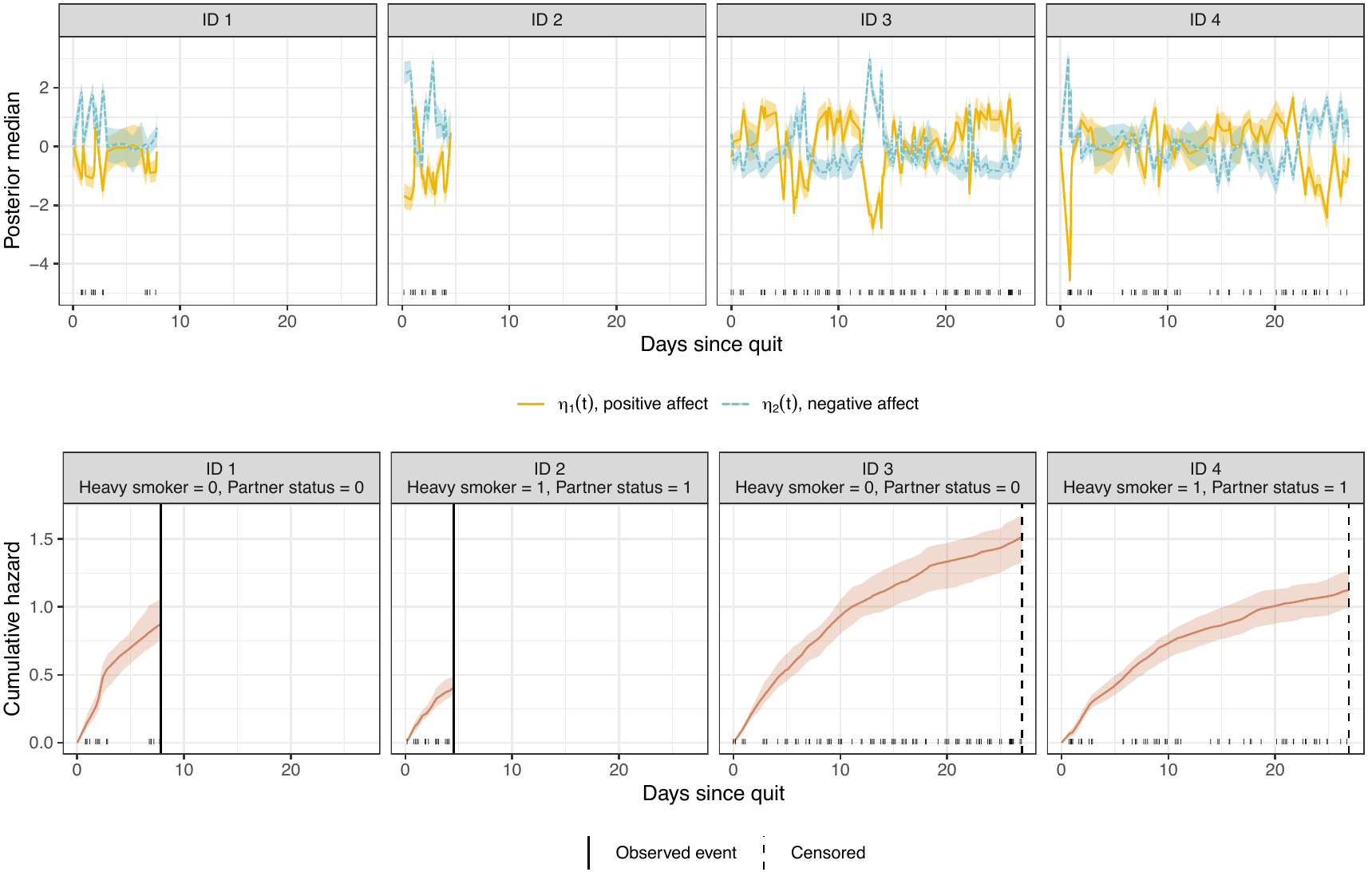}
    \caption{Posterior samples of the latent factors (interpreted as positive and negative affect) and cumulative hazards for four individuals in the mHealth smoking cessation study. Posterior estimates of the latent factors and cumulative hazards are summarized using posterior medians; shaded bands denote the range of values between the 25th and 75th percentiles. The black tick marks along the x-axis of the plots correspond to the longitudinal measurement occasions observed in the data; in the cumulative hazard plots, the vertical lines indicate the timing of observed events or censoring.} \label{fig:care_etas_and_cumul_haz}
\end{figure}

We can also use posterior estimates from our model to examine the trajectory of the latent factors and understand how these latent psychological states of positive and negative affect are linked with the risk of lapse after attempted quit. In Figure \ref{fig:care_etas_and_cumul_haz}, we plot the posterior samples of the two latent factors for positive and negative affect, and the posterior estimates of the cumulative hazard of lapse for four study participants. For periods of follow-up during which the measurement occasions are less frequent, we see increases in the range of values covered by the 25-75\% percentiles of our posterior samples, demonstrating our model's ability to capture the increased uncertainty. We also see the symmetry of our fitted structural submodel reflected in this plot: both positive and negative affect tend to vary in similar ways but in opposite directions. The estimated cumulative hazard functions for these individuals show that the instantaneous risk of lapse is highest immediately after quit time and that the cumulative hazard increases more gradually as time since quit increases.

\section{Discussion}\label{s:discussion}

Motivated by ILD of self-reported emotions collected in an mHealth study of smoking cessation, we propose a joint longitudinal time-to-event model appropriate for modeling ILD. We summarize the multiple longitudinal outcomes as a smaller number of time-varying latent factors using a dynamic factor model with a structure informed by scientific context. These latent factors summarize the multiple longitudinal outcomes (e.g., emotions) and capture vulnerability to an event-time outcome (e.g., risk of lapse). This dimension-reduction approach both simplifies computation and interpretation of the factors associated with altered risk of an event. To fit our model, we use Stan \citep{carpenter_2017}. We integrate out a subset of the latent parameters and leverage a discrete approximation for the survival function to make fitting this model computationally feasible. This proposed approach fills a gap in the literature as a method suitable for modeling multivariate ILD jointly with a time-to-event outcome. While we present models with only two latent factors in this paper, a different (but small) number of factors could also be considered. The choice of number of latent factors could be determined by either domain knowledge or deviance information criterion (DIC). Simulated data and code are available at \texttt{github.com/madelineabbott/OUF\_JM}.

Computational cost is a major concern when jointly fitting a multivariate longitudinal-survival model. In our case, we incorporate a continuous-time stochastic process as a time-varying covariate in our hazard model, increasing the complexity of our likelihood. A Bayesian approach allows us to avoid directly evaluating the complex integrals present in our likelihood, but still requires substantial time due to the repeated sampling within the HMC algorithm. Fitting the joint model in Section \ref{s:data_app} does require about 17 hours (with 4 chains run in parallel on 4 cores) and so further investigation of alternative computational strategies that increase the speed of modeling fitting is an important area of future research. For example, the strategies used in \cite*{murray_2022} and \cite{rustand_2023} could potentially be adapted to work in our setting.

In our approach, we use a discrete approximation of the cumulative hazard function. We show via simulation that simply assuming a somewhat sparse and uniform grid for this approximation works well in the setting of ILD. Furthermore, when ILD is used, our ability to recover good point estimates is not sensitive to the width of the grid. Our specific setting is one in which the ILD captures rapidly varying outcomes and so measurement occasions must occur frequently so that large abrupt fluctuations are not missed. Since measurement occasions are close together, the placement and number of additional grid points are not as important. In other settings in which the longitudinal outcomes change more smoothly, less frequent measurement occasions could still capture important changes over time. When measurement occasions are farther apart, sensitivity to the choice of grid may increase. But, if the longitudinal outcome changes more slowly, linearly interpolating the intermediate values of the longitudinal outcome within the discrete approximation of the hazard function might not be so problematic. Overall, we find that in our ILD setting, the grid width is not particularly important. But one could imagine an alternative scenario where the placement of grid points might matter more. In this scenario, further investigation of the location and number of grid points may be warranted and methods, such as that described in \cite*{fernández_2016}, could be adapted to place grid points according to the intensity of the hazard function. We leave investigation of this alternative scenario as future work.

A weakness of our approach is our assumption of non-informative measurement occasions. Although this assumption is commonly made in joint longitudinal-survival models, self-reported longitudinal outcomes are likely susceptible to informative missingness. Responses to EMA questionnaires may be missing for many reasons, ranging from non-response due to poor mood or lack of cell phone reception. Important future work could include incorporating an additional submodel that accounts for important patterns in the timing of the measurement occasions.

Finally, in our motivating mHealth data, individuals generally experience repeated smoking lapses after attempted quit. We modeled the time until first lapse after attempted quit, but our model could be adapted for the recurrent event of repeated lapses. Additionally, current smokers who attempt to quit often progress through phases in which they are actively attempting to quit before potentially relapsing back into their prior smoking habits. Multi-state models have previously been proposed to model transitions in long-term smoking habits (e.g., \cite{brouwer_2022}). Our joint model could potentially be extended to model short-term changes in cigarette use. Finally, temporal trends could also be incorporated into the longitudinal submodel in order to account for systematic changes over time in the measured longitudinal outcomes, which may vary according to the current state of smoking.

\subsection*{Acknowledgements}

\noindent This work was supported by the National Institutes of Health [grant numbers F31DA057048, R01DA039901, P50DA054039, R01DA014818, P30CA042014, R00CA252604, U01CA229437, U54CA280812] and the Huntsman Cancer Center. The content is solely the responsibility of the authors and does not necessarily represent the official views of the National Institutes of Health or the Huntsman Cancer Foundation. This work is not peer-reviewed.

\subsection*{Conflict of Interest Statement}

\noindent None declared.

\subsection*{Data Availability Statement}

\noindent The mobile health data that support the findings of this study are not publicly available due to privacy restrictions.

\setstretch{1.8}
\bibliographystyle{biometrics}
\bibliography{references}

\setstretch{1.88}

\subsection*{Supporting Information}

\noindent Web Appendices and Figures, referenced in Sections~\ref{s:method}--\ref{s:data_app}, are available with this paper. Example code and simulated data are available on Github at \texttt{github.com/madelineabbott/OUF\_JM}.

\processdelayedfloats

\end{document}


\maketitle

\newpage

\setstretch{1}

\renewcommand{\figurename}{Web Figure}
\renewcommand{\tablename}{Web Table}
\renewcommand{\thesection}{Web Appendix \Alph{section}}

\section{Identifiability and parameter constraints}

\subsection{Converting between OU process parameterizations}

To ensure that our dynamic factor model is identifiable, we model the Ornstein-Uhlenbeck (OU) process on the correlation scale.  A $p$-dimensional OU process is generally parameterized in terms of two $p$-dimensional OU process, $\theta$ and $\sigma$, where $\theta$ controls the speed of mean reversion and $\sigma$ controls the volatility of the process.  The stochastic differential equation (SDE) form of a bivariate OU process, denoted by $\eta_1$ and $\eta_2$, is:

$$d \begin{bmatrix} \eta_{1}(t) \\ \eta_{2}(t) \end{bmatrix} =  - \underbrace{\begin{bmatrix} \theta_{11} & \theta_{12} \\ \theta_{21} & \theta_{22} \end{bmatrix}}_{\coloneqq \bm{\theta}_{OU}} \begin{bmatrix} \eta_{1}(t) \\ \eta_{2}(t) \end{bmatrix} dt + \underbrace{\begin{bmatrix} \sigma_{11} & \sigma_{21} \\ \sigma_{12} & \sigma_{22} \end{bmatrix}}_{\coloneqq \bm{\sigma}_{OU}} d\begin{bmatrix} W_{1}(t) \\ W_{2}(t) \end{bmatrix}$$

\noindent where $W_1(t)$, $W_2(t)$ are standard Brownian motion.  The OU process is most often written in terms of its conditional distribution.  Assuming that the initial value of the latent process is $\bm{\eta}(t_0) \sim N_p(\bm{0}, \bm{V})$, where $\bm{V} = vec^{-1}\big\{(\bm{\theta}_{OU} \oplus \bm{\theta}_{OU})^{-1} vec(\bm{\sigma}_{OU} \bm{\sigma}_{OU}^{\top}) \big\}$, then for $j = 1, ...$,

\begin{align*}
    \bm{\eta}(t_{j}) | \bm{\eta}(t_{j-1}) \sim N_p \left(e^{-\bm{\theta}_{OU} (t_{j} - t_{j-1})} \bm{\eta}(t_{j-1}), \bm{V} - e^{- \bm{\theta}_{OU} (t_{j} - t_{j-1})} \bm{V} e^{- \bm{\theta}^{\top}_{OU} (t_{j} - t_{j-1})}\right)
\end{align*}

If the OU process is modeled on the correlation scale, then $\bm{V}$ is a correlation matrix.  We use $\rho$ to denote the off-diagonal parameter(s) of $\bm{V}$.  For a bivariate OU process, we must estimate only one off-diagonal parameter in $\bm{V}$, where $$\bm{V} = \begin{bmatrix} 1 & \rho \\ \rho & 1 \end{bmatrix}.$$

When estimating the parameters that describe the OU process, we could do so using the parameterization in the SDE or the parameterization of the conditional distribution.  That is, we could estimate either $(\bm{\theta}_{OU}, \bm{\sigma}_{OU})$ or $(\bm{\theta}_{OU}, \rho)$.

While the SDE version of the OU process is a function of $\bm{\sigma}_{OU}$, the conditional distribution is a function of $\bm{\sigma}_{OU} \bm{\sigma}_{OU}^{\top}$ (that is, $\bm{\sigma}_{OU}$ never shows up outside of the term $\bm{\sigma}_{OU} \bm{\sigma}_{OU}^{\top}$).  We know that $\bm{\sigma}_{OU} \bm{\sigma}_{OU}^{\top}$ is symmetric and positive definite.  Because  $\bm{\sigma}_{OU} \bm{\sigma}_{OU}^{\top}$ is symmetric, this means that the conditional distribution only depends on the identifiable parameter $\bm{\sigma}_{OU} \bm{\sigma}_{OU}^{\top}$ (plus additional parameters in $\bm{\theta}_{OU}$).  In the case of the bivariate OU process, $\bm{\sigma}_{OU} \bm{\sigma}_{OU}^{\top}$ contains three identifiable parameters (that is, the upper \textit{or} lower triangle is identifiable).

Although $\bm{\sigma}_{OU}$ is not immediately identifiable from a known $\bm{\theta}_{OU}$ and $\bm{V}$, we can always re-parameterized an arbitrary OU process with a full $\bm{\sigma}_{OU}$ to have a triangular $\bm{\sigma}_{OU}$ while maintaining the same correlation structure.  This fact results from the stationary covariance formula being a function of $\bm{\sigma}_{OU} \bm{\sigma}_{OU}^{\top}$:

$$ \bm{V} = vec^{-1}\{(\bm{\theta}_{OU} \oplus \bm{\theta}_{OU})^{-1} vec[\bm{\sigma}_{OU} \bm{\sigma}_{OU}^{\top}]\} $$

\noindent We can decompose  $\bm{\sigma}_{OU} \bm{\sigma}_{OU}^{\top}$ into $\bm{\sigma}_{OU}$ using the Cholesky decomposition, which says that if matrix $A$ is positive definite, then $\bm{A}$ can be decomposed into two lower-triangular matrices $\bm{L}$, where $\bm{A} = \bm{L} \bm{L}^{\top}$.  Furthermore, since $\bm{\sigma}_{OU} \bm{\sigma}_{OU}^{\top}$ is positive definite, then we know that the Cholesky decomposition of $\bm{\sigma}_{OU} \bm{\sigma}_{OU}^{\top}$ is unique.  To convert an OU process with parameters $\bm{\theta}^*_{OU}, \rho$ to parameters $\bm{\theta_{OU}}, \bm{\sigma}_{OU}$, we can simply solve the stationary variance $\bm{V}$ (containing parameter(s) $\rho$) for $\bm{\sigma}_{OU} \bm{\sigma}_{OU}^{\top}$.

\subsection{Constraints on $\theta_{OU}$}

\cite{tran_2021} discuss constraints on the OU process $\bm{\theta}_{OU}$ matrix, which ensure that $\bm{\theta}_{OU}$ corresponds to a mean reverting process but does not restrict it from being non-oscillating. We apply these constraints developed in this prior work, which constrain the real parts of the eigenvalues of bivariate OU process parameter $\bm{\theta}_{OU}$ to be positive:

\begin{align*}
    v_1 &= \theta_{OU_{11}} + \theta_{OU_{22}} \\
    v_2 &= \theta_{OU_{11}} \theta_{OU_{22}} - \theta_{OU_{12}}\theta_{OU_{21}}
\end{align*}

\noindent where $v_1$ and $v_2$ must be positive.  \cite{tran_2021} also discuss eigenvalue constraints for a trivariate OU process.

\section{Simulation study design}

\subsection{Measurement patterns}

In our simulation study, we generate observations of our longitudinal outcomes using four different patterns in measurement occasions:

\begin{itemize}
    \item[] \textbf{Measurement pattern 1}: The measurements occur frequently and with constant probability.  To determine the timing of the measurements, we sample uniformly from a fine grid of possible times spanning 0 to 28 days, where a maximum of 60 and 70 measurement times are drawn in settings 1 and 2, respectively.  After censoring of the longitudinal measurements due to the survival outcome, an average of 19.2 and 24.4 measurement occasions are observed per individual in settings 1 and 2, respectively.
    \item[] \textbf{Measurement pattern 2}: The measurements occur less frequently but still with constant probability.  To determine the timing of the measurements, we sample uniformly from a fine grid of possible times spanning 0 to 28 days, where a maximum of 15 and 12 measurement times are drawn in settings 1 and 2, respectively.  After censoring of the longitudinal measurements due to the survival outcome, an average of 5.5 and 5.0 measurement occasions are observed per individual in settings 1 and 2, respectively.
    \item[] \textbf{Measurement pattern 3}: The measurements are distributed according to the measurement times observed in the motivating mHealth study, CARE.  To determine the timing of the measurements, we sample (with replacement) individuals from the motivating mHealth dataset and then use their observed measurement times to define the measurement times in the simulated dataset. After censoring of the longitudinal measurements due to the survival outcome, an average of 33.5 and 36.9 measurement occasions are observed per individual in settings 1 and 2, respectively.
    \item[] \textbf{Measurement pattern 4}: The measurements are clustered together and distributed according to probabilities following a truncated cosine function of time.  To determine the timing of the measurements, we generate measurement probabilities for each point on the fine grid of possible measurement times going from 0 to 28 days.  These measurement probabilities are proportional to the absolute value of a cosine function and, in order to induce larger gaps between clusters of measurements, are truncated to 0 when less than 0.4.  After censoring of the longitudinal measurements due to the survival outcome, an average of 29.3 and 20.4 measurement occasions are observed per individual in settings 1 and 2, respectively.
\end{itemize}

\begin{figure}
    \centering
    \captionsetup{width=\linewidth}
    \includegraphics[width=\linewidth]{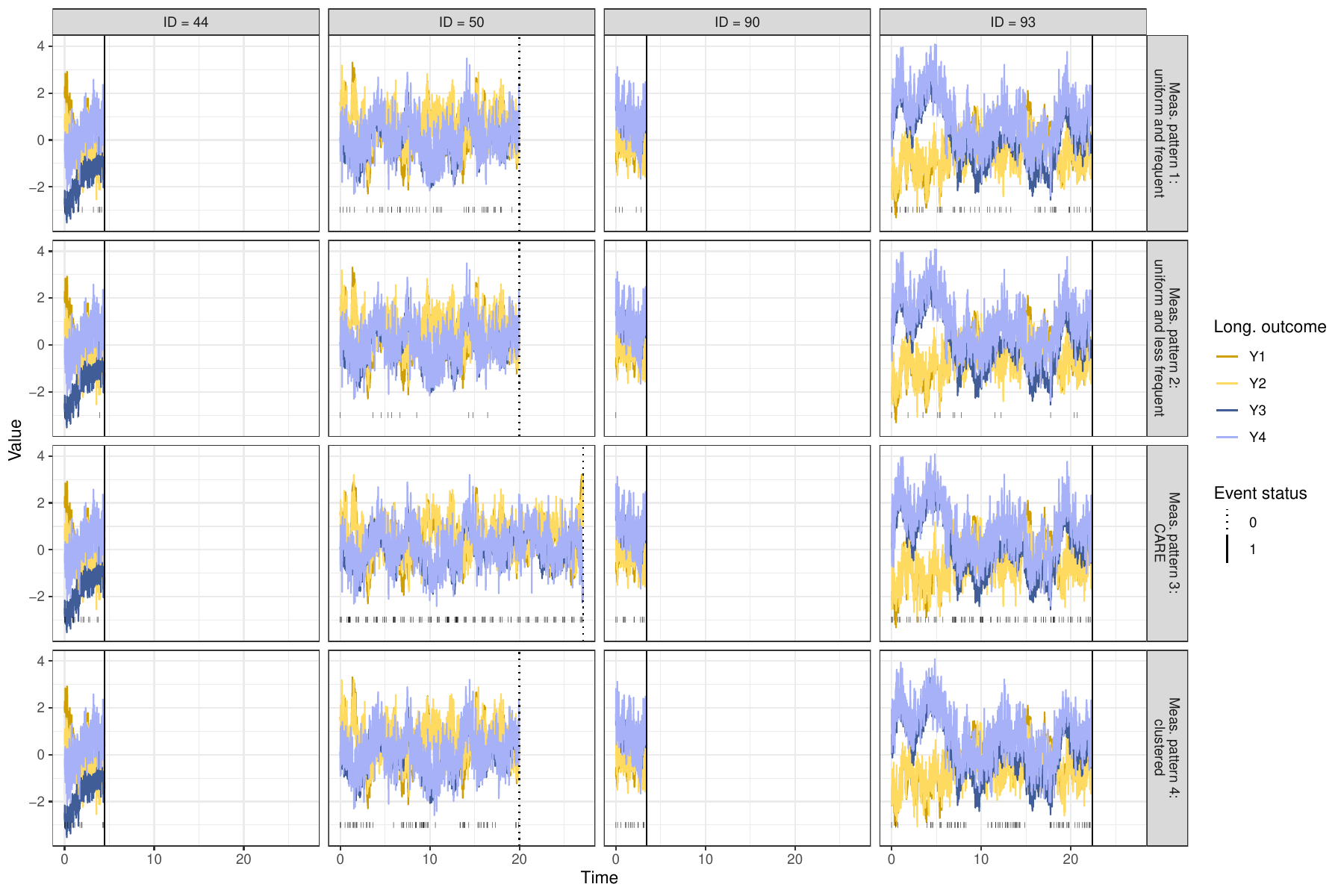}
    \caption{Each column displays the true values of the four longitudinal outcomes for a different individual in the simulated data under \textbf{setting 1}.  Each row displays the different patterns in measurement occasions, where the timing of the measurement occasions are indicated by the location of the black tick marks along the x-axis.} \label{supp_fig:meas_occ_s1}
\end{figure}

\begin{figure}
    \centering
    \captionsetup{width=\linewidth}
    \includegraphics[width=\linewidth]{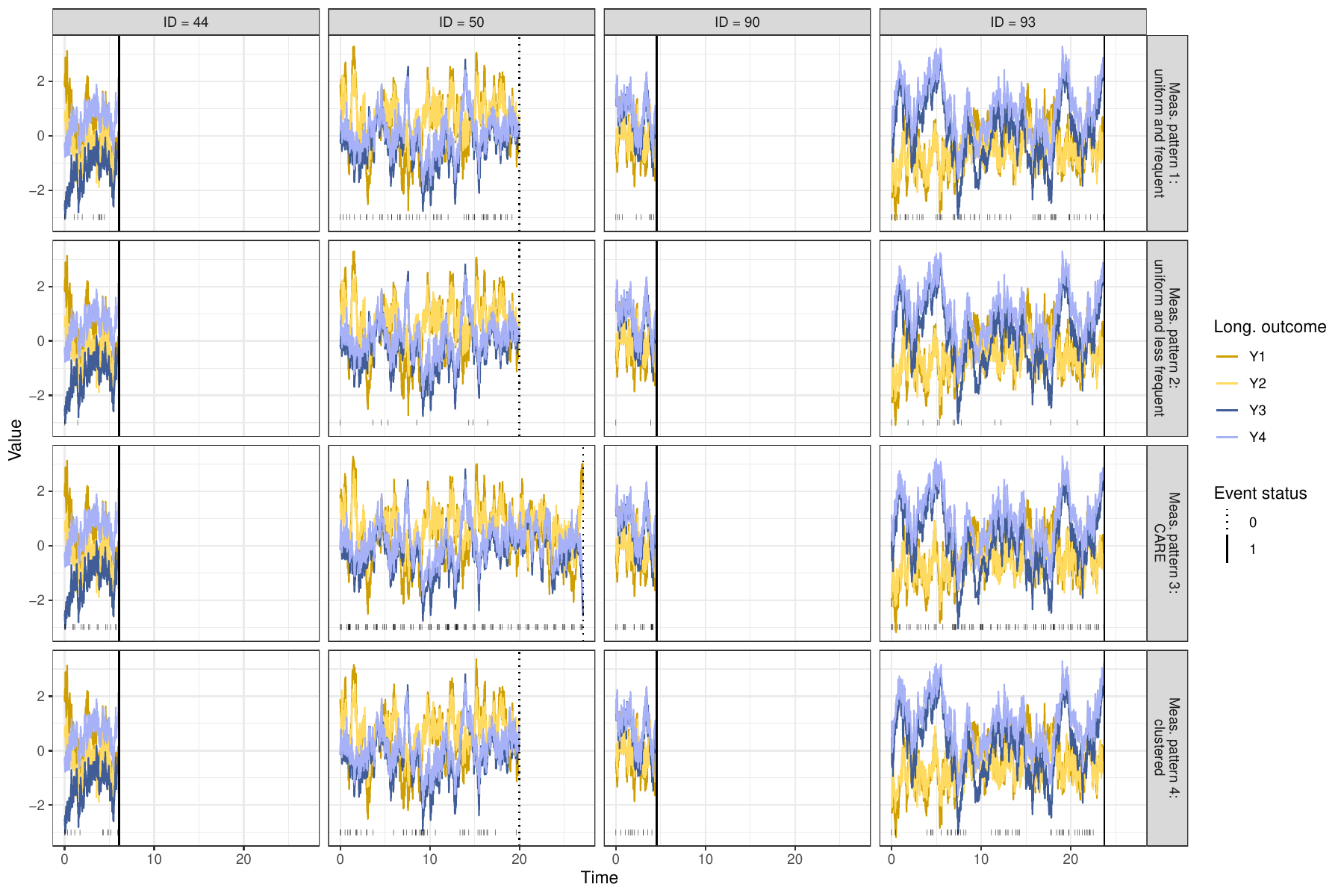}
    \caption{Each column displays the true values of the four longitudinal outcomes for a different individual in the simulated data under \textbf{setting 2}.  Each row displays the different patterns in measurement occasions, where the timing of the measurement occasions are indicated by the location of the black tick marks along the x-axis.} \label{supp_fig:meas_occ_s2}
\end{figure}

We illustrate the different patterns in measurements in Web Figures \ref{supp_fig:meas_occ_s1} and \ref{supp_fig:meas_occ_s2} for four individuals in datasets simulated under setting 1 and setting 2.  Only measurements that occurred prior to the event or censoring time are displayed in this figure.

\newpage

\subsection{True parameter values}

In setting 1, we assume the following true parameter values:

Measurement submodel: $$\bm{\theta}_{OU} = \begin{bmatrix} 1.8 & 0.4 \\ 1.5 & 1.2 \end{bmatrix} \text{,  } \bm{\sigma}_{OU} = \begin{bmatrix} 1.76 & 0\\ 0 & 0.71 \end{bmatrix} \implies \rho = -0.633$$

Structural submodel: $$\bm{\Lambda} = \begin{bmatrix} 0.9 & 0 \\ 0.5 & 0 \\ 0 & 1 \\ 0 & 0.8 \end{bmatrix}$$
    $$ \sigma_{u_1} = 0.4, \sigma_{u_2} = 0.5, \sigma_{u_3} = 0.8, \sigma_{u_4} = 1$$
    $$\sigma_{\epsilon_1} = 0.2, \sigma_{\epsilon_2} = 0.6, \sigma_{\epsilon_3} = 0.3, \sigma_{\epsilon_4} = 0.7$$

Survival submodel: $$\beta_0 = -2.5, \beta_1 = -0.2, \beta_2 = 0.3$$

\noindent In setting 2, we assume the following true parameters values:

Measurement submodel: $$\bm{\theta}_{OU} = \begin{bmatrix} 2.4 & 0.4 \\ 0.8 & 2 \end{bmatrix} \text{,  } \bm{\sigma}_{OU} = \begin{bmatrix} 2.14 & 0\\ 0 & 1.89 \end{bmatrix} \implies \rho = -0.273$$

Structural submodel: $$\bm{\Lambda} = \begin{bmatrix} 0.9 & 0 \\ 0.5 & 0 \\ 0 & 1 \\ 0 & 0.8 \end{bmatrix}$$
    $$ \sigma_{u_1} = 0.4, \sigma_{u_2} = 0.5, \sigma_{u_3} = 0.8, \sigma_{u_4} = 1$$
    $$\sigma_{\epsilon_1} = 0.2, \sigma_{\epsilon_2} = 0.3, \sigma_{\epsilon_3} = 0.1, \sigma_{\epsilon_4} = 0.2$$

Survival submodel: $$\beta_0 = -3, \beta_1 = -0.4, \beta_2 = 0.8$$

\vspace{0.5cm}

\begin{figure}
    \centering
    \captionsetup{width=\linewidth}
    \includegraphics[width=0.7\linewidth]{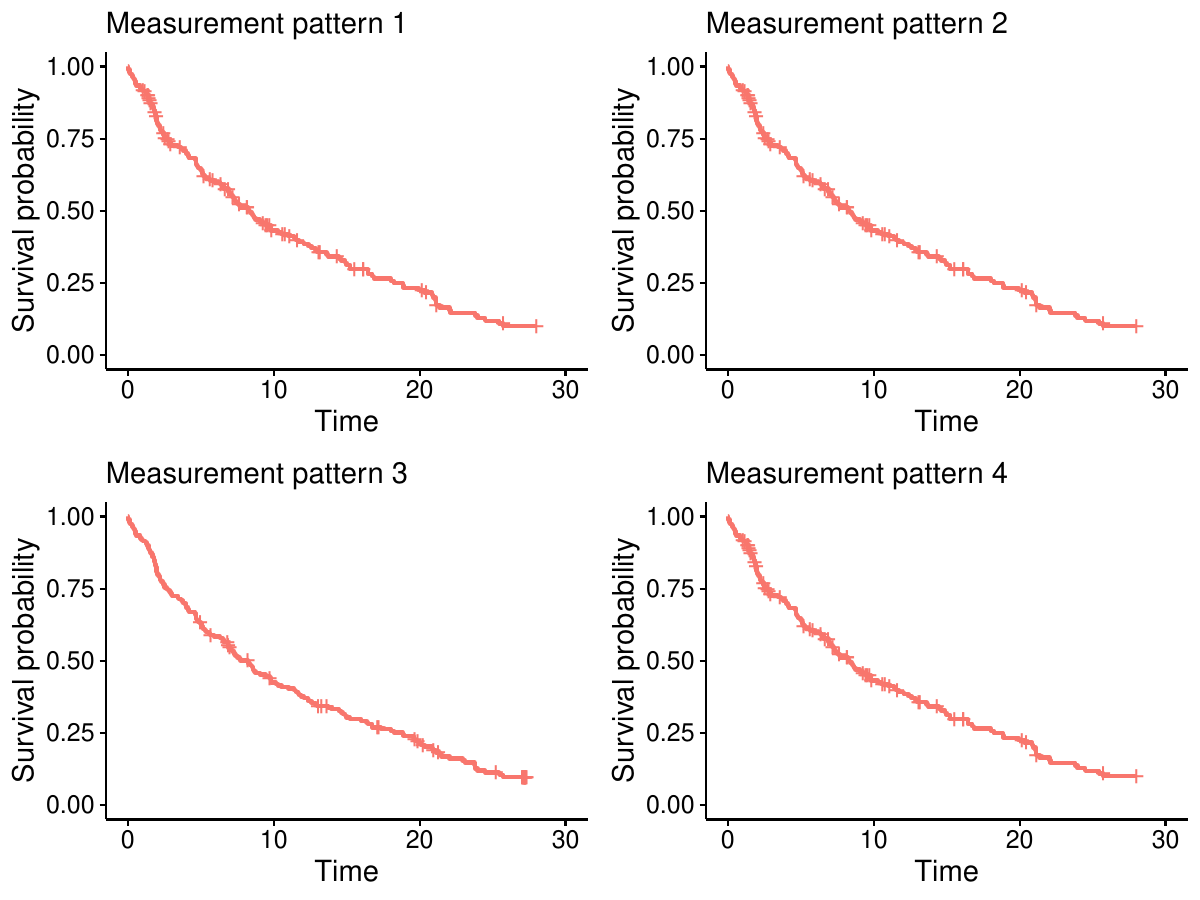}
    \caption{Kaplan-Meier curves for a single dataset with each measurement pattern (1-4) simulated using the true set of parameters in \textbf{setting 1}.} \label{supp_fig:km_sim_setting1}
\end{figure}

\begin{figure}
    \centering
    \captionsetup{width=\linewidth}
    \includegraphics[width=0.7\linewidth]{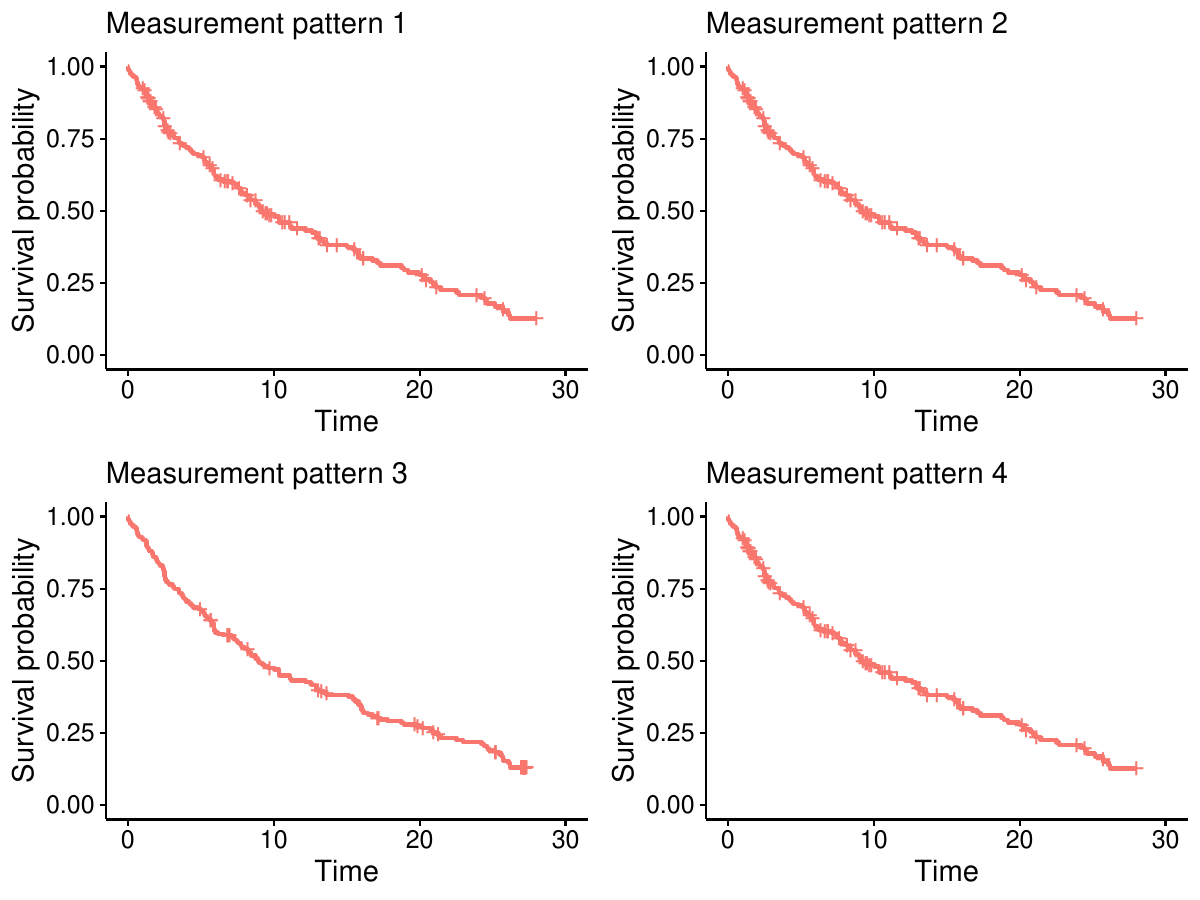}
    \caption{Kaplan-Meier curves for a single dataset with each measurement pattern (1-4) simulated using the true set of parameters in \textbf{setting 2}.} \label{supp_fig:km_sim_setting2}
\end{figure}

In both settings 1 and 2, the true survival submodel is $h_i(t) = exp(\beta_0 + \beta_1 \eta_{1i}(t) + \beta_2 \eta_{2i}(t))$.  The true censoring distribution is $C_i \sim 10 \times Exponential(rate = 0.25)$ for measurement patterns 1, 2, and 4.  For measurement pattern 3 in which the timing of measurements is based on those observed in the motivating mHealth study, the timing of an individual's final random ecological momentary assessment (EMA) is used as the censoring time.  Kaplan-Meier curves are plotted in Web Figures \ref{supp_fig:km_sim_setting1} and \ref{supp_fig:km_sim_setting2} for each combination of setting and measurement pattern.

\subsection{Initial parameter values}

When fitting the model in the simulation study (both settings 1 and 2), we specify initial parameter values that are correct in their sign and in their order of magnitude.  The specific values are listed below.  For parameters not listed below (e.g., $\eta$), we rely on the random initial values generated by Stan \citep{carpenter_2017}.  In practice, we could take a two-stage approach to initialization.

Measurement submodel: $$\bm{\theta}_{OU} = \begin{bmatrix} 1 & 0.5 \\ 0.5 & 1 \end{bmatrix} \text{,  } \rho = -0.5$$

Structural submodel: $$\bm{\Lambda} = \begin{bmatrix} 1 & 0 \\ 1 & 0 \\ 0 & 1 \\ 0 & 1 \end{bmatrix}$$
    $$ \sigma_{u_k} = 0.1, k = 1, ..., 4$$
    $$\sigma_{\epsilon_k} = 0.1, k = 1, ..., 4$$

Survival submodel: $$\beta_0 = -1, \beta_1 = -1, \beta_2 = 1$$

\subsection{Prior distributions}

We specify the following prior distributions when fitting the models in our simulation study.  These priors are based on those used in \cite{tran_2021}.

\begin{align*}
    &\lambda_k \sim \text{half-}N(1, \sigma^2_\lambda); k = 1, ..., 4 \\
    &\sigma_\lambda \sim \text{half-Cauchy}(0, 5) \\
    &\theta_{OU_{11}}, \theta_{OU_{21}}, \theta_{OU_{12}}, \theta_{OU_{22}} \sim N(0, 10^2) \\
    &\rho \sim \text{Uniform}(-0.999999, 0.999999) \\
    &\sigma_{u_k} \sim \text{half-}Cauchy(0, 5); k = 1, ..., 9 \\
    &\sigma_{u_{\epsilon}} \sim \text{half-}Cauchy(0, 5); k = 1, ..., 9 \\
    &\beta_0, \beta_1, \beta_2 \sim N(0, 5^2)
\end{align*}

\newpage

\section{Investigation of grid width}

\begin{figure}[H]
    \centering
    \captionsetup{width=\linewidth}
    \includegraphics[width=\linewidth]{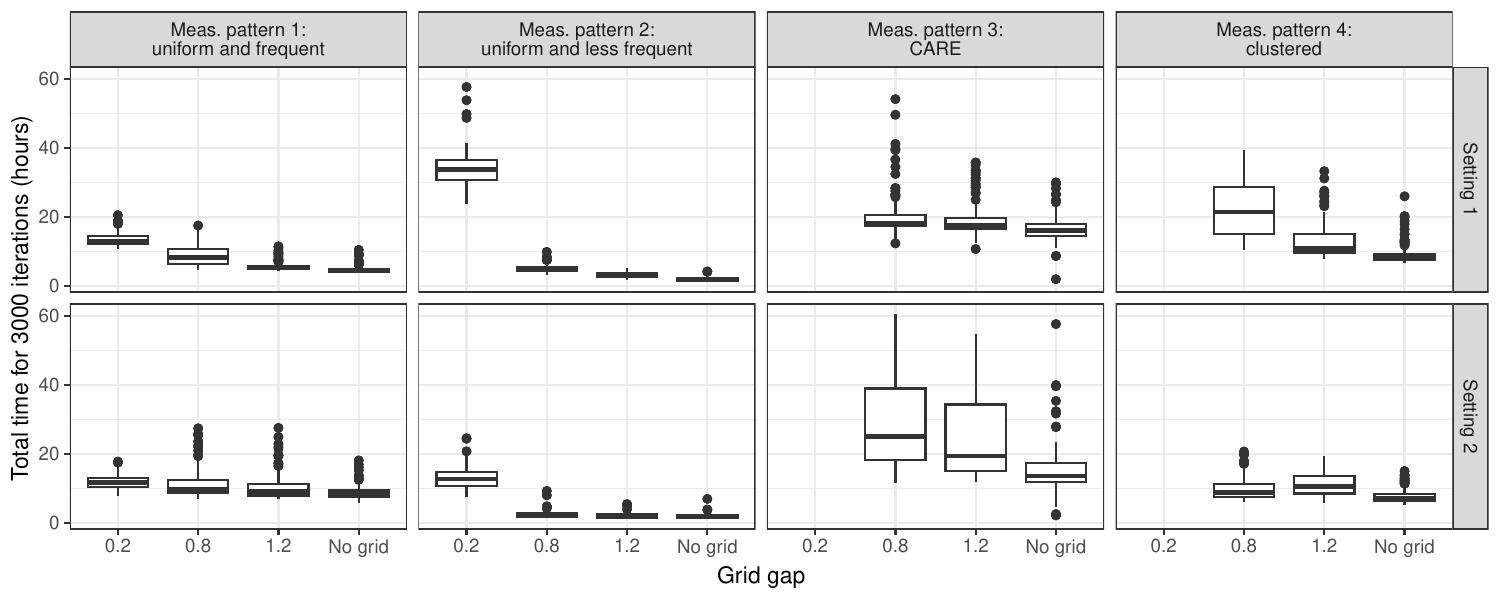}
    \caption{For data generated under settings 1 and 2 with each of the four measurement patterns, we summarize the time required to run Stan for 3000 iterations.} \label{supp_fig:comp_time}
\end{figure}

For the simulation study described in the main text, we present complete results in this section for all combinations of true parameter values (settings 1 and 2), measurement patterns (1-4), and grid widths (0.2, 0.4, 1.2, and no grid).  Due to the high computation cost, we do not fit models using the finest grid (0.2) for data generated under measurement patterns 3 and 4.  Computation times are summarized in Web Figure \ref{supp_fig:comp_time}.

\vspace{1cm}

\vspace{1cm}

\begin{figure}
    \centering
    \captionsetup{width=\linewidth}
    \includegraphics[width=\linewidth]{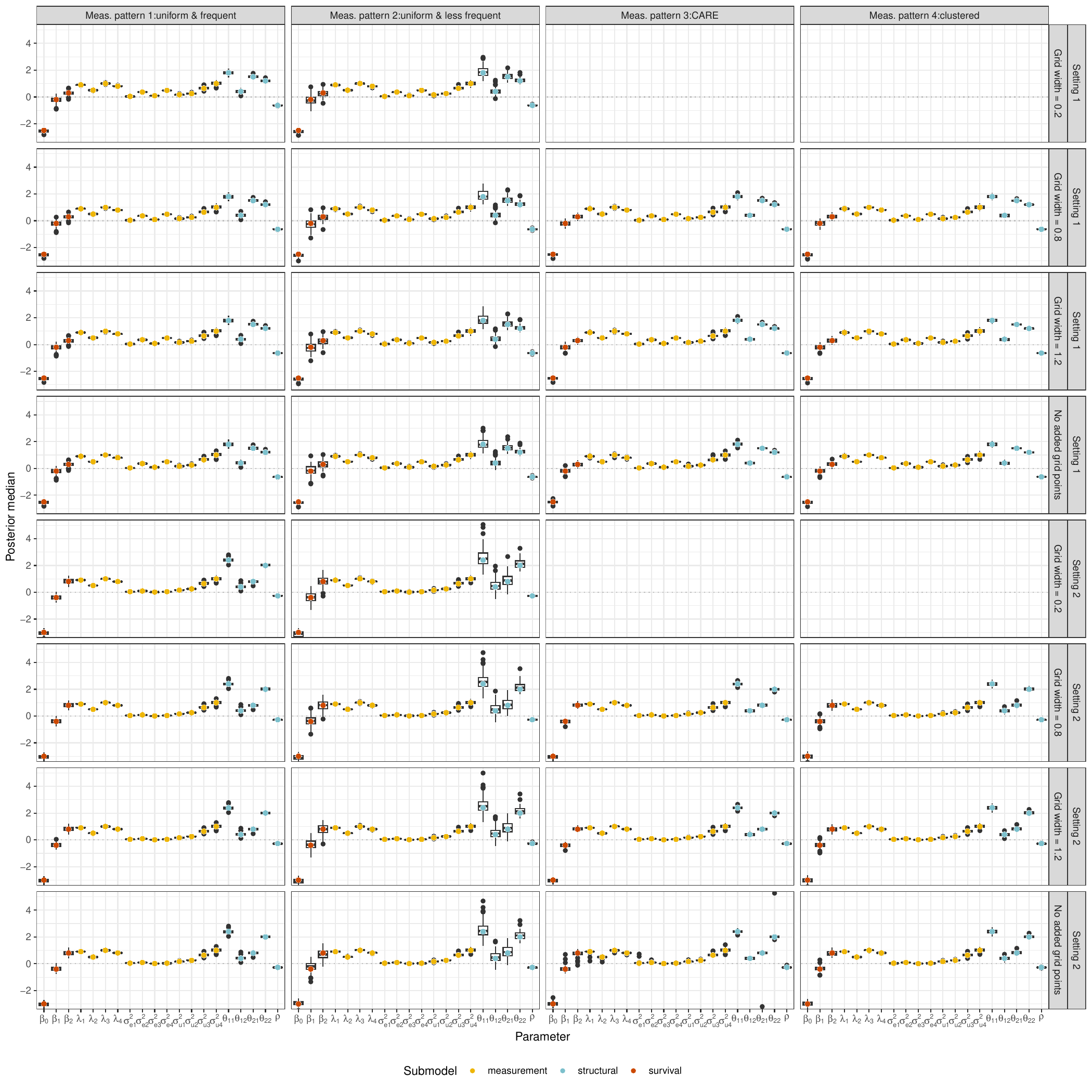}
    \caption{For data generated under settings 1 and 2 with each of the four measurement patterns, we use box plots to summarize the distribution of the \textbf{posterior medians for all parameters} across the 100 simulated datasets.  True parameter values are indicated with colored dots. Note that the y-axis range is truncated to span -3 to 5 and so some posterior medians corresponding to measurement pattern 3 and no added grid points are not shown; results for this combination of measurement pattern and grid width are re-plotted separately in Web Figure 7.} \label{supp_fig:post_medians_grid}
\end{figure}

\begin{figure}
    \centering
    \captionsetup{width=\linewidth}
    \includegraphics[width=0.7\linewidth]{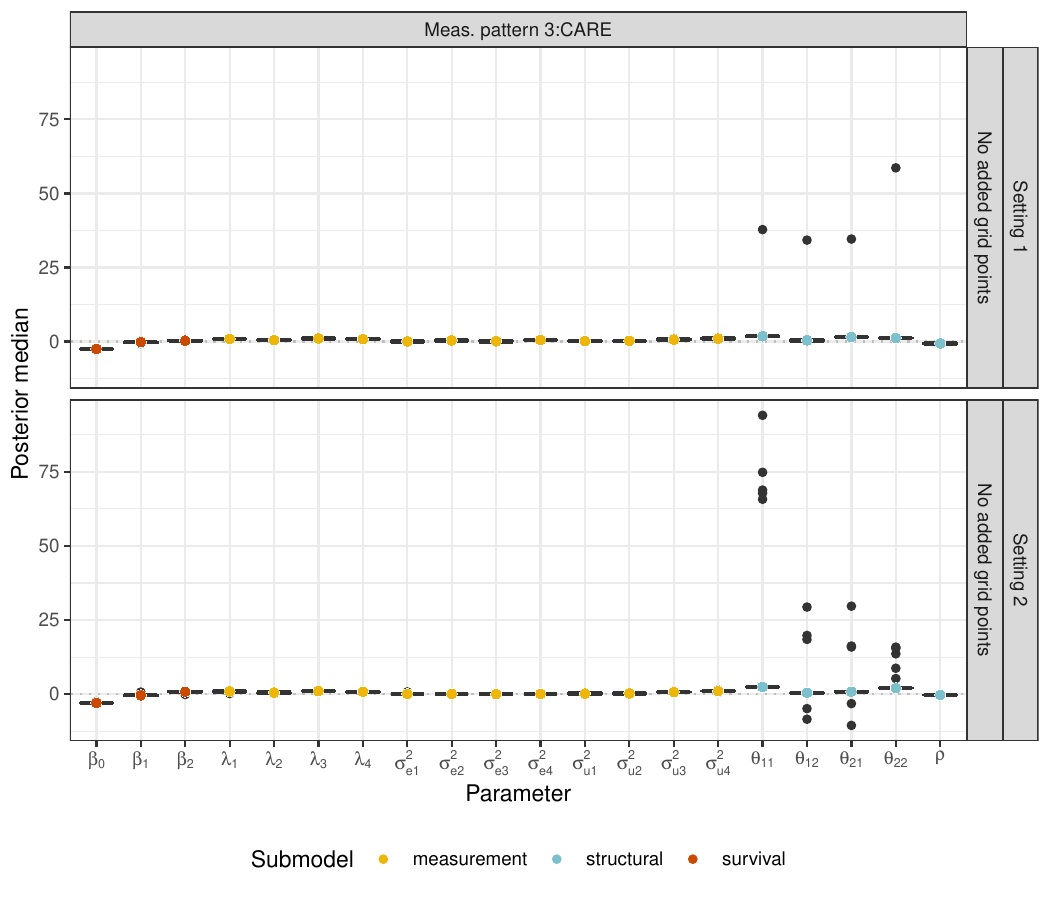}
    \caption{For data generated under settings 1 and 2 with \textbf{measurement pattern 3}, we use box plots to summarize the distribution of the \textbf{posterior medians for all parameters} across the 100 simulated datasets when fitting the model \textbf{without added grid points}.  True parameter values are indicated with colored dots.  These plots show the same set of results as in Web Figure 6 for measurement pattern 3 with no additional grid points, but here we allow a wider range of values on the y-axis so that all posterior medians are visible in the plot.} \label{supp_fig:post_medians_grid2}
\end{figure}

Posterior medians are summarized by grid width in Web Figure \ref{supp_fig:post_medians_grid}.  In some cases, when we fit the joint model with no added grid points to data generated under measurement pattern 3, we encounter issues with convergence and the posterior medians take extreme values.  Due to the range of the y-axis in Web Figure \ref{supp_fig:post_medians_grid}, the instances in which posterior medians take extreme values do not appear in this figure.  As such, we re-plot the results for this scenario (measurement pattern 3, no added grid points) separately in Web Figure \ref{supp_fig:post_medians_grid2}.  For both setting 1 and 2 with measurement pattern 3, we find that adding grid points resolves this issue; the posterior medians are close to the truth when we fit the model using a grid of width 0.8 or 1.2, vs. no grid at all.

\begin{figure}
    \centering
    \captionsetup{width=\linewidth}
    \includegraphics[width=\linewidth]{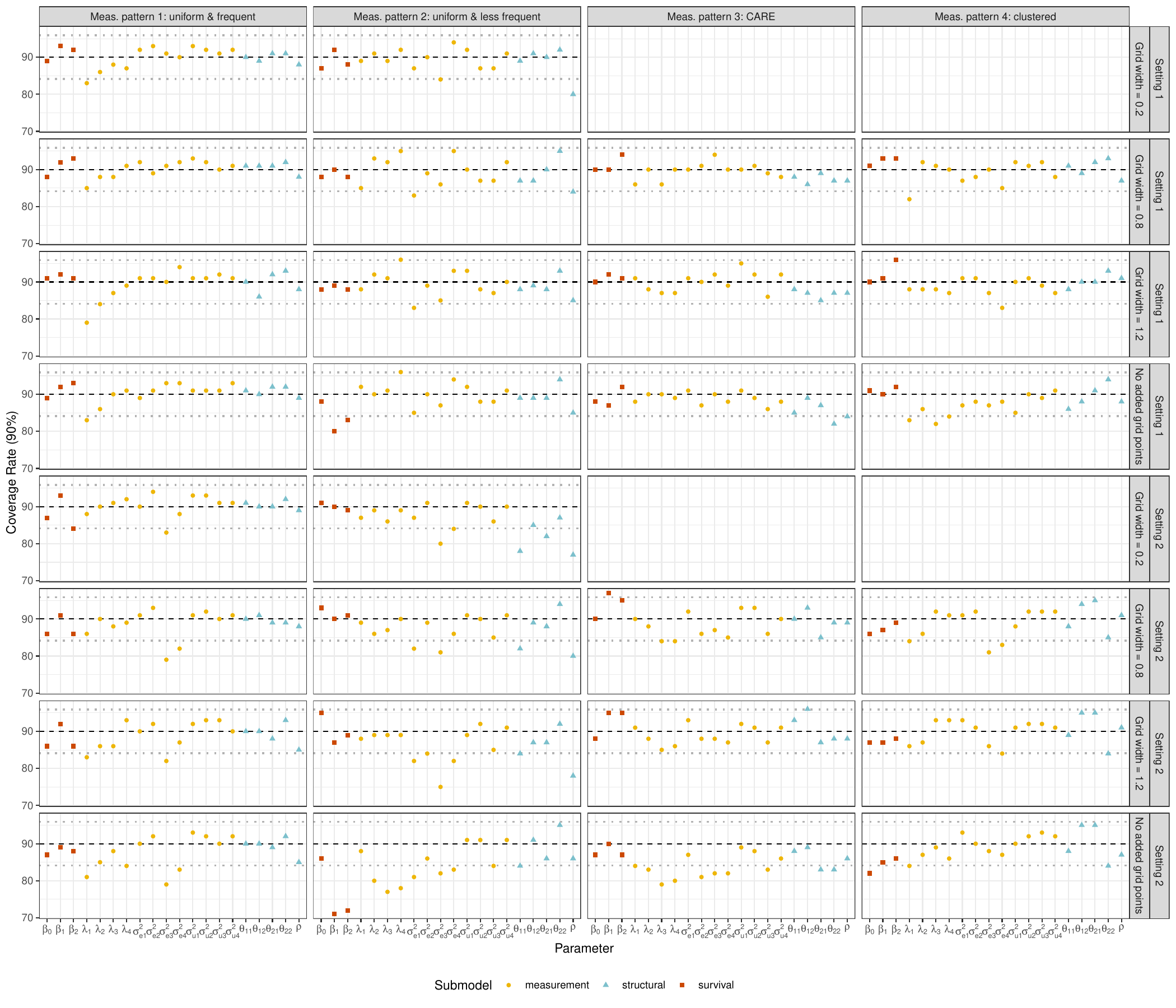}
    \caption{For data generated under settings 1 and 2 with each of the four measurement patterns, we summarize the \textbf{coverage rate of 90\% credible intervals} across the 100 simulated datasets with the colored dots. The black dashed line indicates target coverage and the grey dashed lines mark the expected range of values based on a 90\% binomial proportion confidence interval.} \label{supp_fig:cov_rates}
\end{figure}

Coverage rates are summarized by grid width in Web Figure \ref{supp_fig:cov_rates}.

\newpage

\section{Analysis of data from smoking cessation study}

Our analytic sample consists of 238 individuals who did not lapse within the first 12 hours of the post-quit period (i.e., within 12 hours of 4am on the quit day) and who also responded to the emotion-related questions in at least one random EMA after the first 12 hours of the study.  We also required that the baseline covariates of pre-quit smoking history and partner status be non-missing.

\paragraph{Definition of the survival outcome:} In \cite{vinci_2017}, the authors analyze the same dataset to assess associations between position emotions and smoking habits after attempted quit. The authors conduct two analyses: (a) they assess the association between pre-quit positive emotions and a binary outcome of lapse on the first day after quit and (b) the association between post-quit positive emotions and the risk of first lapse (as a time-to-event outcome) among individuals who did not lapse on the first day after quit. In the first analysis, the authors used data from all individuals; in the second analysis, the authors restricted their final dataset to the subset of individuals who did not lapse on the first day.

\begin{figure}
    \centering
    \captionsetup{width=\linewidth}
    \includegraphics[width=0.75\linewidth]{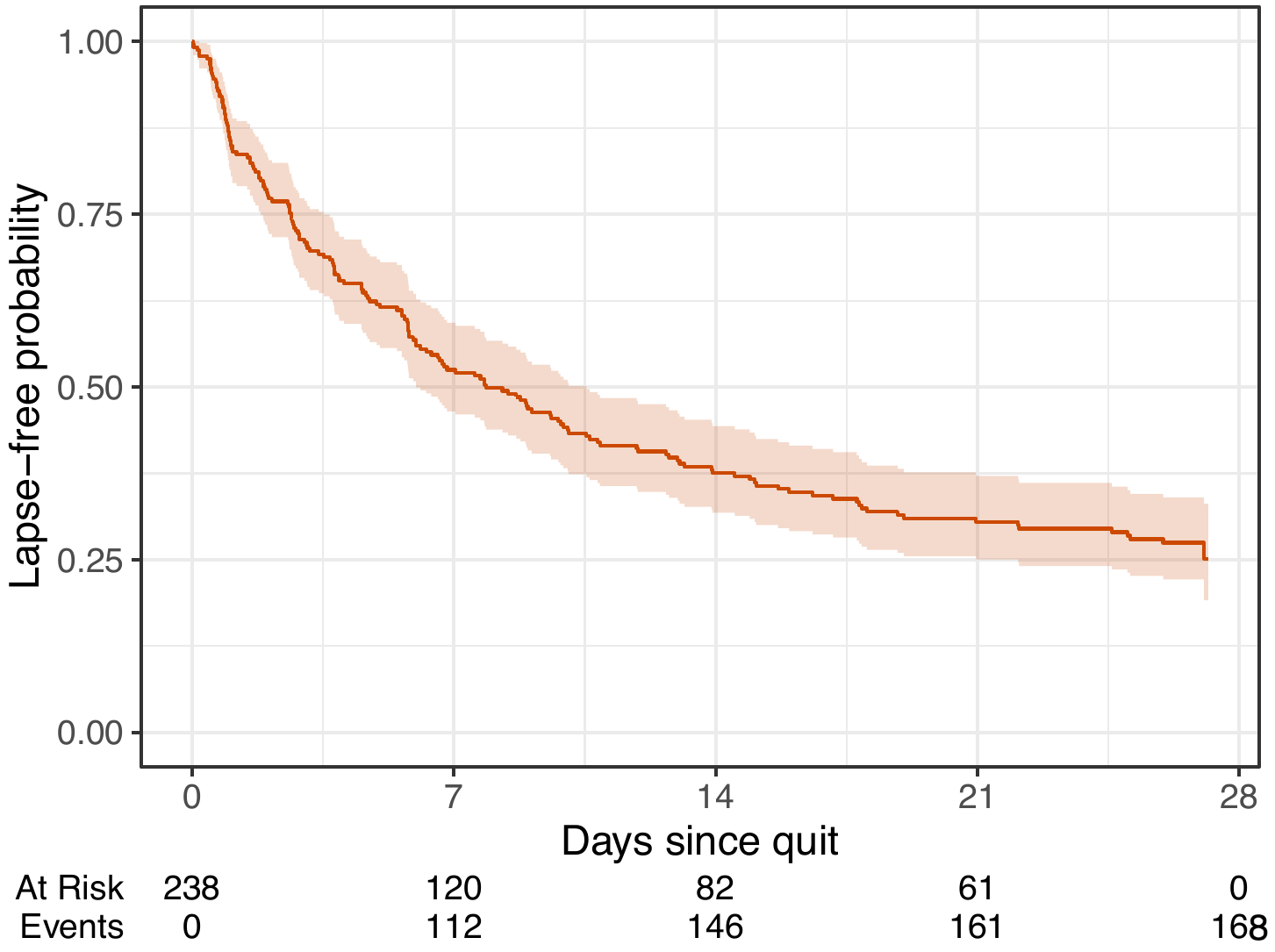}
    \caption{The Kaplan-Meier curve for the time-to-event outcome of time until lapse. The time origin corresponds to 4pm on the reported quit day.} \label{supp_fig:km_lapse}
\end{figure}

We take the same subsetting approach here in order to avoid uncertainty surrounding the exact time of quit and reduce the number of lapse events that occur within minutes of the assumed quit time. We opt to use 12 hours as our cutoff, rather than 24, since we assume that the day of quit is known even if the exact time is not.  A Kaplan-Meier curve for time-to-first-lapse for those who did not lapse within the first 12 hours is given in Web Figure \ref{supp_fig:km_lapse}.  Given that individuals who lapse almost immediately contribute limited information when fitting the model, we do not expect our results to be particularly sensitive to our exact definition of the time origin (e.g., 4pm (which we use) vs. 5pm vs. 7pm, for example). Sensitivity analyses could be conducted to better assess the impact of our assumed quit time on the fitted model.

\newpage

\subsection{Specification of piecewise constant baseline hazard}

\cite{lazaro_2021} describe a regularized version of a piecewise constant baseline hazard where the priors are specified such that the segments are correlated.  The baseline hazard, made up of B segments, is 

\begin{align*}
    h_0(t ; \beta_0) = \sum_{i = 1}^{B} \beta_{0_b} \bm{I}(c_{b-1} < t \leq c_b)
\end{align*}

\noindent where $\beta_0 = (\beta_{0_1}, ..., \beta_{0_B})$, $\bm{I}$ is an indicator function, $c_0 = 0$, and $c_B$ is the time of the final (observed or censored) event time.  We assume $B = 10$ and that the segments are of equal length.  Then, prior distributions are specified as

\begin{align*}
    p(log(\beta_{0_1})) &\sim N(0, \sigma_{\beta}^2) \\
    p(log(\beta_{0_2})) &\sim N(log(\beta_{0_1}), \sigma_{\beta}^2) \\
    &\vdots \\
    p(log(\beta_{0_B})) &\sim N(log(\beta_{0_{B-1}}), \sigma_{\beta}^2)
\end{align*}

\noindent where $\sigma^2_{\beta} \sim \text{half-Cauchy}(0, 25)$, as suggested in \cite{gelman_2006}.

\subsection{Prior distributions}

We report prior distributions for the remainder of the parameters here:

\begin{align*}
    &\lambda_k \sim \text{half-}N(1, \sigma^2_\lambda); k = 1, ..., 9 \\
    &\sigma_\lambda \sim \text{half-Cauchy}(0, 5) \\
    &\theta_{OU_{11}}, \theta_{OU_{21}}, \theta_{OU_{12}}, \theta_{OU_{22}} \sim N(0, 10^2) \\
    &\rho \sim \text{Uniform}(-0.999999, 0.999999) \\
    &\sigma_{u_k} \sim \text{half-}Cauchy(0, 5); k = 1, ..., 9 \\
    &\sigma_{u_{\epsilon}} \sim \text{half-}Cauchy(0, 5); k = 1, ..., 9 \\
    &\beta_1, \beta_2 \sim N(0, 5^2) \\
    &\alpha_1, \alpha_2 \sim N(0, 5^2)
\end{align*}

\subsection{Model diagnostic plots}

\begin{figure}
    \centering
    \captionsetup{width=\linewidth}
    \includegraphics[width=\linewidth]{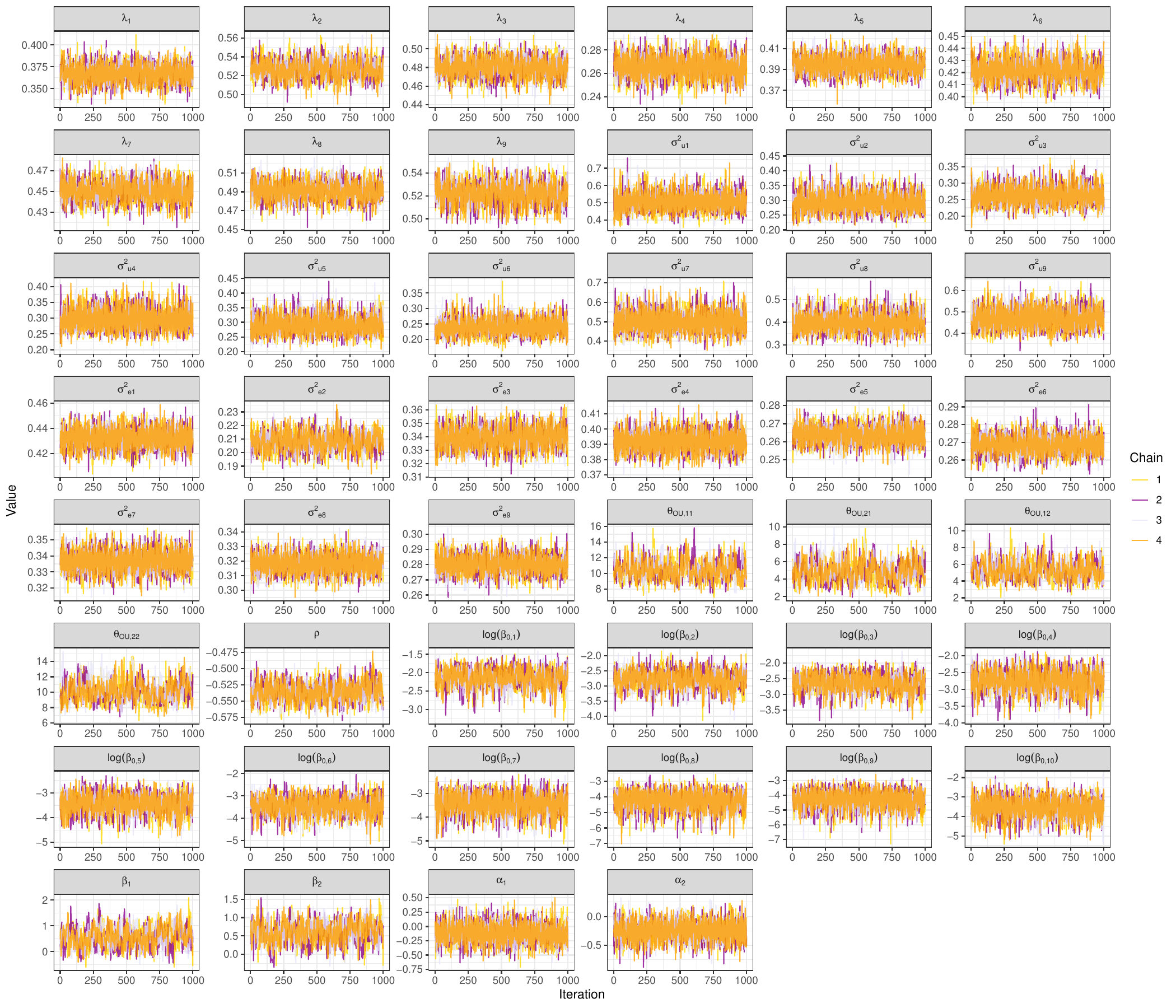}
    \caption{Trace plot of posterior samples (after burn-in) for the joint model with the \textbf{piecewise constant baseline hazard} fit to the mHealth smoking cessation study.} \label{webfig:traceplot_h0pc}
\end{figure}

\begin{figure}
    \centering
    \captionsetup{width=\linewidth}
    \includegraphics[width=\linewidth]{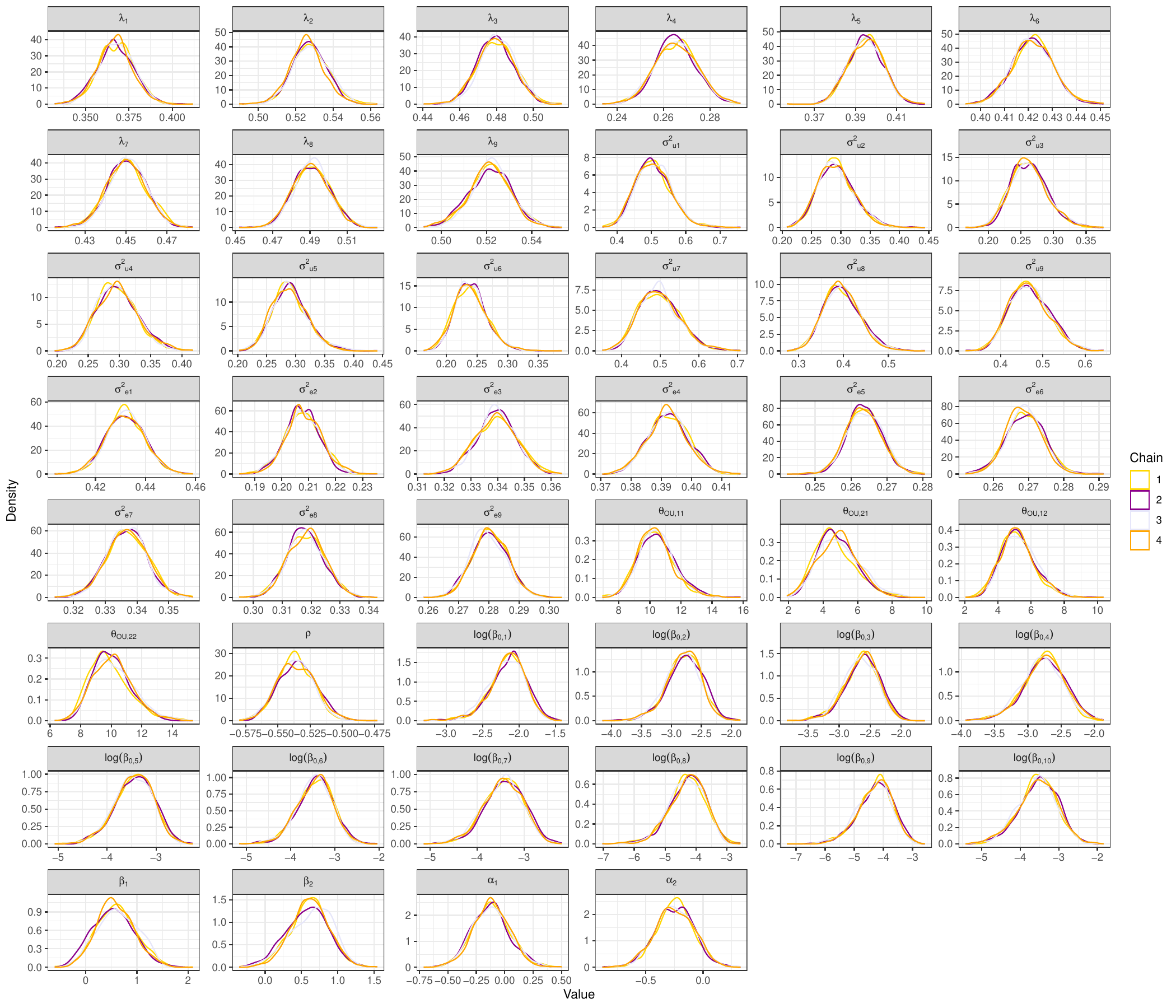}
    \caption{Posterior densities of parameters for the joint model with the \textbf{piecewise constant baseline hazard} applied to the mHealth smoking cessation study.} \label{webfig:postdens_h0pc}
\end{figure}

\begin{figure}
    \centering
    \captionsetup{width=\linewidth}
    \includegraphics[width=\linewidth]{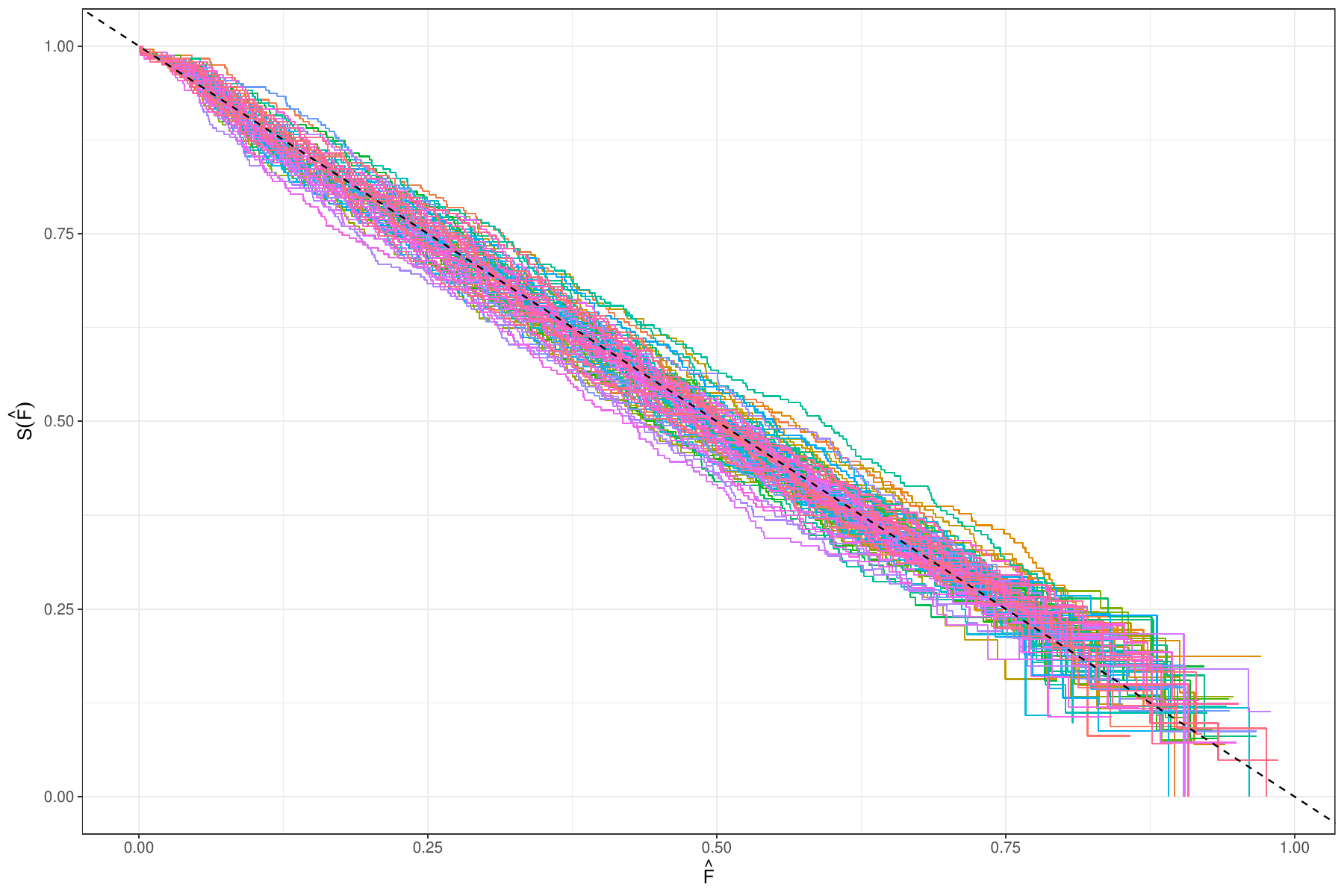}
    \caption{Goodness of fit for the joint model with the \textbf{piecewise constant baseline hazard} in the survival submodel.  Each solid line corresponds to the Kaplan-Meier survival curve calculated from a single set of posterior samples; curves are plotted for 100/1000 total posterior samples samples.  If the model fits well, then we expect the Kaplan-Meier curves to follow the dashed line going from $(1,1)$ to $(1,0)$.} \label{webfig:surv_goodness_of_fit_h0pc}
\end{figure}

For the joint model with the piecewise constant baseline hazard, we provide trace plots and posterior densities in Web Figures \ref{webfig:traceplot_h0pc} and \ref{webfig:postdens_h0pc}.  We also provide a plot that evaluates the goodness-of-fit of our model in Web Figure \ref{webfig:surv_goodness_of_fit_h0pc}; this plot shows the distribution of the predicted survival probabilities.  This approach to assessing goodness-of-fit uses a strategy similar to Cox-Snell residuals: We know that if $T \sim S(t)$ and $F(t) = 1 - S(t)$, then $F(T) \sim Unif(0,1)$.  So, for each set of posterior samples for $\hat{\beta}$ and $\hat{\eta}$ for all $i = 1, ..., N$, we:
\begin{enumerate}
    \item Calculate $\hat{S}_i(T_i)$ using $\hat{\beta}$ and $\hat{\eta}$ where $T_i$ is the observed event time
    \item Fit a Kaplan-Meier curve to $(1-\hat{S}_i(T_i), \delta_i)$
\end{enumerate}

\noindent If our predicted survival probabilities are accurate, they should be approximately uniformly distributed between 0 and 1 and so the Kaplan-Meier curve should follow a diagonal line from $(1,1)$ to $(1,0)$.

\begin{figure}
    \centering
    \captionsetup{width=\linewidth}
    \includegraphics[width=\linewidth]{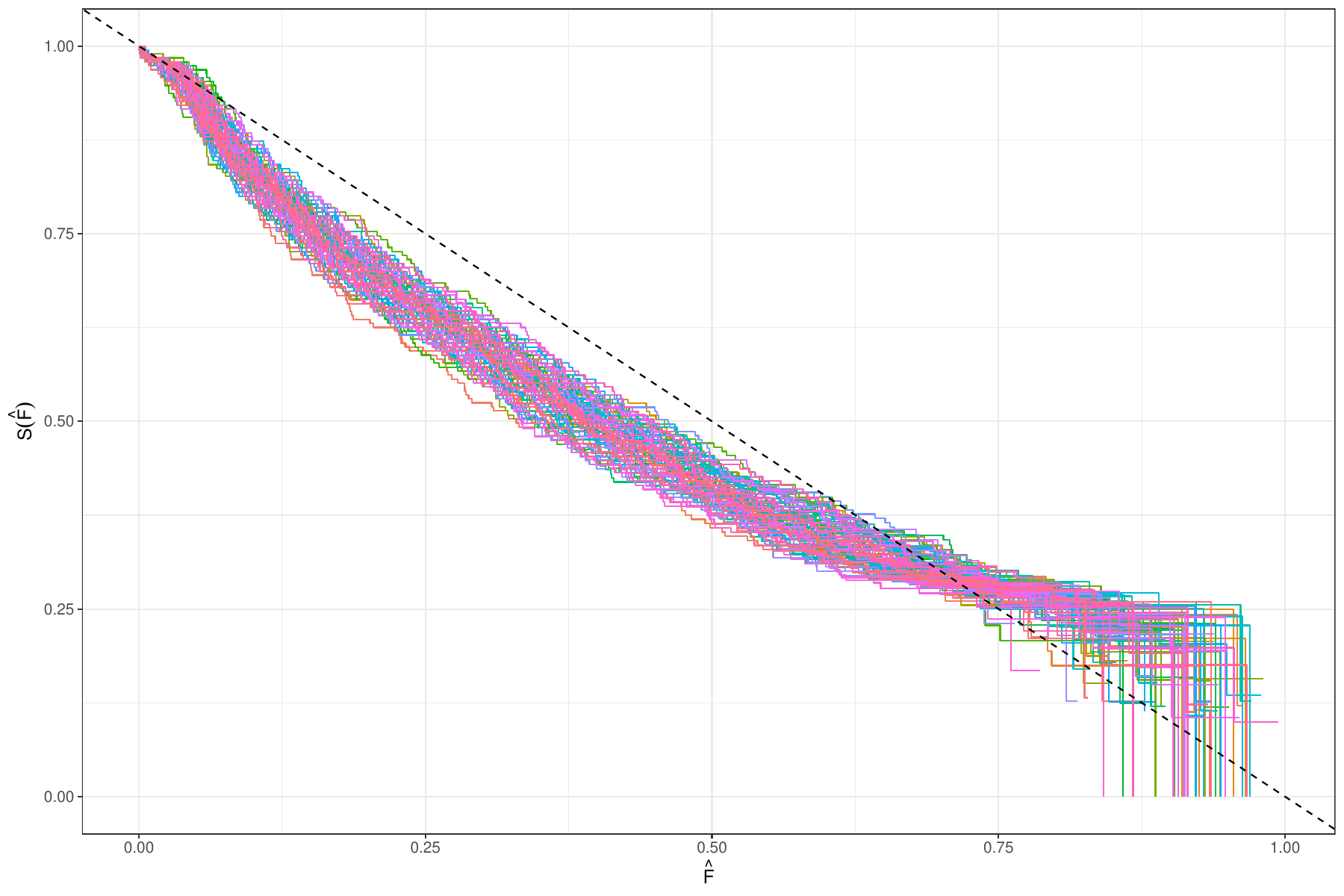}
    \caption{Goodness of fit for the joint model with the \textbf{Weibull baseline hazard} in the survival submodel.  Each solid line corresponds to the Kaplan-Meier survival curve calculated from a single set of posterior samples; curves are plotted for 100/1000 total posterior samples samples.  If the model fits well, then we expect the Kaplan-Meier curves to follow the dashed line going from $(1,1)$ to $(1,0)$.} \label{webfig:surv_goodness_of_fit_h0weibull}
\end{figure}

We compared the model with the piecewise constant baseline hazard to a model with a Weibull baseline hazard function and found that the piecewise constant baseline hazard appeared to result in a model that better fit the data.  The goodness-of-fit plot for the fitted Weibull model is given in Web Figure \ref{webfig:surv_goodness_of_fit_h0weibull}.

\newpage 

\subsection{Plotting correlation decay in the latent factors}

When estimating the OU process, we rely on the conditional distribution.  To plot the estimated correlation decay in the latent factors across increasing time intervals (as in Figure 5 in the main manuscript), we use the (unconditional) covariance formula.  We provide this marginal covariance formula below.  Assuming that $\eta(s)$ and $\eta(t)$ are two observation of latent factors from an OU process at times $s$ and $t$, where $s < t$, then the marginal joint distribution is:

$$\begin{bmatrix} \bm{\eta}(s) \\ \bm{\eta}(t) \end{bmatrix} \sim N \left(\begin{bmatrix} 0 \\ 0 \end{bmatrix}, \bm{\Psi} = \begin{bmatrix} \bm{V} & \bm{V} e^{-\bm{\theta}^{\top}(t - s)} \\ e^{-\bm{\theta}(t - s)} \bm{V} & \bm{V} \end{bmatrix} \right)$$

To calculate the estimated correlation between the latent factors across increasing time intervals, we plug posterior samples of the structural submodel parameters into the off-diagonal elements of $\bm{\Psi}$ along with $s = 0$ and increasing values of $t$.  Because our identifiability constraint assumes that we model the OU process on the correlation scale, the covariance matrix above will be the correlation matrix here. In Figure 5 in the main manuscript, the x-axis corresponds to increasing values of $t$ and the y-axis corresponds to different elements of $\bm{\Psi}$.

\newpage

\bibliographystyle{biometrics}
\bibliography{references}
